\documentclass[12pt,preprint]{aastex}

\def\dnu{$\Delta\nu$}
\def\numax{$\nu_{\rm max}$}
\def\teff{$T_{\rm eff}$}
\def\logg{$\log g$}

\shorttitle{Grid based asteroseismology}
\shortauthors{Gai et al.}

\begin{document}

\title{An in-depth study of grid-based asteroseismic analysis}

\author{Ning Gai\altaffilmark{1,2}, Sarbani Basu\altaffilmark{2},
William J. Chaplin\altaffilmark{3}, Yvonne Elsworth\altaffilmark{3}}

\altaffiltext{1}{Department of Astronomy, Beijing Normal University,
Beijing 100875, China; gaining@mail.bnu.edu.cn}

\altaffiltext{2}{Department of Astronomy, Yale University, P.O. Box
208101, New Haven, CT 06520-8101, USA}
\email{sarbani.basu@yale.edu}

\altaffiltext{3}{School of Physics and Astronomy, University of
Birmingham, Edgbaston, Birmingham B15 2TT, UK; w.j.chaplin@bham.ac.uk,
y.p.elsworth@bham.ac.uk}

\begin{abstract}

NASA's {\it Kepler} mission is providing basic asteroseismic data for
hundreds of stars. One of the more common ways of determining stellar
characteristics from these data is by so-called ``grid based''
modelling. We have made a detailed study of grid-based analysis
techniques to study the errors (and error-correlations) involved. As
had been reported earlier, we find that it is relatively easy to get
very precise values of stellar radii using grid-based
techniques. However, we find that there are small, but significant, biases that can
result because of the grid of models used. The biases can be minimized
if metallicity is known. Masses cannot be determined as precisely as
the radii, and suffer from larger systematic effects. We also find
that the errors in mass and radius are correlated. A positive consequence of
this correlation is that \logg\ can be determined both precisely and
accurately with almost no systematic biases. Radii and \logg\ can be
determined with almost no model dependence to within 5\% for realistic
estimates of errors in asteroseismic and conventional observations.
Errors in mass can be somewhat higher unless accurate metallicity
estimates are available.  Age estimates of individual stars are the
most model dependent.  The errors are larger too. However, we find
that for star-clusters, it is possible to get a relatively precise age
if one assumes that all stars in a given cluster have the same age.

\end{abstract}

\keywords{methods: data analysis -- stars: fundamental parameters --
stars: interiors -- stars: oscillations}

\section{Introduction}

NASA's {\it Kepler} mission (Borucki et al. 2010) is observing
solar-like oscillations in hundreds of stars. While we expect to get
individual frequencies in a good fraction of these stars, the initial
data are the so-called large frequency separation, \dnu, and the
frequency of maximum oscillations power, \numax. These data, combined
with conventional observations of effective temperature, \teff, and
metallicity, [Fe/H], are being used to determine the basic properties
of the stars.

In order to analyze {\it Kepler} data on a large number of stars, a
variety of semi-automated and fully automated pipelines have been
developed. These pipelines are based on seismic and non-seismic
properties of precomputed grids of stellar models. Characteristics of
stars are determined by searching among the models to get a ``best
fit'' for a given observed set of (\dnu, \numax, \teff, and
[Fe/H]). This is usually referred to as ``grid'' asteroseismology.
Different groups define their ``best fits'' in different ways.  Stello
et al. (2009) and Basu et al. (2010) have described the
radius-determination pipeline of various groups. While all pipelines
have been tested to some extent, there have been no large-scale tests
to determine systematic errors in the results and the effect that the
underlying grid of models might have on the results.  Additionally,
although these pipelines were constructed primarily to determine
stellar radii, they have been modified to determine other stellar
parameters such as mass, \logg\ and age (see e.g., Metcalfe et
al. 2010), however, there have been no tests to determine the errors
involved in determining these parameters. In this paper we rectify
this oversight and test different aspects of grid asteroseismology.

We use three grids of models to test the model-dependence of grid
asteroseismology results. We use the Yale-Birmingham pipeline (Basu et
al. 2010; described in \S~\ref{subsec:grid}) as the basis. 
We also estimate mass and radius directly from
\dnu, \numax, and \teff\ to determine whether or not grid methods are
really needed to estimate these quantities. We test whether errors in
the estimated mass and radius are correlated. And although determining
ages of single stars is difficult (and model dependent), we examine
whether seismic data allow us to do better than conventional fitting
of evolutionary tracks.

This paper is organized as follows: we describe the analysis
technique, including the grids of models, in \S~\ref{sec:method}.  Our
results are discussed in \S~\ref{sec:res}, where we present results
for radius, mass, \logg\ and age estimations.  We present our
conclusions in \S~\ref{sec:conc}.

\section {Method}
\label{sec:method}

\subsection{Direct Method}
\label{subsec:dir}

 The direct method of determining stellar radii and masses depends on
the availability of data on  \dnu, 
\numax, and the effective temperature \teff.

For the solar-type stars, oscillation spectra present patterns of
peaks that show nearly regular separations in frequency. The large
frequency separation, \dnu, is the most obvious of these separations
and is the spacing between consecutive overtones of the same spherical
angular degree, $l$. When the signal-to-noise ratios in the seismic
data are insufficient to allow robust extraction of individual
oscillation frequencies, it is still possible to extract estimates of
the average large frequency separation for use as the seismic input
data. In fact this is the case for many {\it Kepler} stars. The
average large separation is formally related to the mean density of a
star (see e.g., Christensen-Dalsgaard 1993). Large separations can be
calculated as
 \begin{equation}
 \frac{\Delta\nu}{\Delta\nu_{\odot}}=\sqrt{\frac{M/M_{\odot}}{(R/R_{\odot})^{3}}},
 \label{eq:delnu}
 \end{equation}
assuming we know the \dnu\ for the Sun. Stello et al. (2009) have shown
that this scaling holds over most of the HR diagram and errors
are probably below 1\%.

The frequency of maximum power in the oscillations power spectrum,
\numax, is related to the acoustic cut-off frequency of a star
(e.g., see Kjeldsen \& Bedding 1995; Bedding \& Kjeldsen 2003;
Chaplin et al. 2008), which in turn scales as $M\,R^{-2}\,T_{\rm
eff}^{-1/2}$. Thus, if we know the solar value of \numax, we can
calculate \numax\ for any star as:
 \begin{equation}
 {{\nu_{\rm max}}\over{\nu_{{\rm max},\odot}}}={{M/M_{\odot}}\over
 {(R/R_{\odot})^2\sqrt{(T_{\rm eff}/T_{{\rm eff},\odot})}}}
 \label{eq:numax}
 \end{equation}
If \dnu, \numax\ and \teff are known, Equations~(\ref{eq:delnu}) and
(\ref{eq:numax}) represent two equations in two unknowns $M$ and $R$
and hence, can be solved to obtain both $M$ and $R$. This also
allows us to calculate \logg.

\subsection{The Grid Method}
\label{subsec:grid}

Basu et al.~(2010) described the Yale-Birmingham (YB) pipeline to
determine stellar radii. 
 Briefly, the
pipeline is based on finding the maximum likelihood of the set of
input parameters calculated with respect to the grid models.
For a given observational (central) input parameter set, the first key step
in the method is generating 10,000 input parameter sets by adding
different random realizations of Gaussian noise to the actual
(central) observational input parameter set. The distribution of
radii obtained from the central parameter set
and the 10,000 perturbed parameter sets form the distribution
function. 
 The final estimate of the parameter is the median of the distribution.
Basu et al.~(2010) used the inter-quartile distance of the distribution
function as a measure of the errors in radius.

The results presented here are based on the YB pipeline with small
modifications. Basu et al. (2010) used models that were part of the
Yale-Yonsei isochrones (Demarque et al. 2004) as their grid. Here, we
use three different grids that are described below in
\S~\ref{subsec:mod}. Basu et al.(2010) used the center of gravity of
the likelihood function with respect to radius as the estimate of
radius for each of the 10001 sets of inputs. We found that while that
works well for radius, better estimates of mass, age, \logg\
\emph{and} radius, are obtained if we take an average of the points
that have the highest likelihood. We average all points with
likelihoods over 95\% of the maximum value of the likelihood
functions. Additionally, instead of using quartile points to define
the error, we use 1$\sigma$ limits (i.e. the lower 34\% and the upper
34\%) as a measure of the errors, to be consistent with other groups
that do grid modelling.

We use a combination of \dnu, \numax, \teff\ and [Fe/H] as inputs.
The likelihood function is formally defined as
 \begin{equation}
 \mathcal {L}=(\prod^{n}_{i=1}\frac{1}{\sqrt{2\pi}\sigma_{i}})\times
 \exp(-\chi^{2}/2), \label{eq:likelihood}
 \end{equation}
where
 \begin{equation}
 \chi^{2}=\sum^{n}_{i=1}(\frac{q^{obs}_{i}-q^{model}_{i}}{\sigma^{i}})^{2},
 \label{eq:chi2}
 \end{equation}
with $q$ $\equiv$ \{$T_{\rm eff}$, [Fe/H], $\Delta\nu$, $\nu_{\rm max}$\} and
$\sigma$ are the nominal errors in input parameters.  From the form of
the likelihood function in Equation (\ref{eq:likelihood}) it is
apparent that we can easily include more inputs, or drop some
inputs. For determining ages of stars, we also use the grid method
without using any seismic data, but use the absolute visual magnitude,
$M_V$, instead.

\subsection{Databases used}
\label{subsec:mod}

Our main grid of models is the YREC grid. These are models that we
constructed using the Yale Rotation and Evolution Code (YREC; Demarque
et al.~2008) in its non-rotating configuration. The input physics
includes the OPAL equation of state tables of Rogers \& Nayfonov
(2002), and OPAL high temperature opacities (Iglesias \& Rogers~1996)
supplemented with low temperature opacities from Ferguson et
al.~(2005). The NACRE nuclear reaction rates (Angulo et al.~1999) were
used. All models included gravitational settling of helium and heavy
elements using the formulation of Thoul et al.~(1994).  We use the
Eddington $T$-$\tau$ relation, and the adopted
mixing length parameter is $\alpha$ $\equiv$ 1.826. An overshoot of
$\alpha_{c}$ = 0.2$Hp$ was assumed for models with convective cores.

The model grid has about 820,000 individual models. These models have
$[Fe/H]$ ranging from +0.6 to $-0.6$ dex in steps of 0.05 dex. We
assume that [Fe/H]=0 corresponds to the solar abundance ($Z/X=0.023$)
as determined by Grevesse \& Sauval (1998) and these models have a
helium abundance of $Y=0.246$ (i.e., the observed solar helium
abundance; Basu \& Antia 2008). The helium abundance for other models
with other values of metallicity was determined assuming a chemical
evolution model $\Delta Y/\Delta Z$ = 1. For each [Fe/H], we have
models with $M$ = 0.80 to 3.0 $M_{\odot}$ and the spacing in mass is
0.02 $M_{\odot}$. The age of the models were restricted between 0.02
and 15 Gyr. When needed, we used the color tables of Lejeune et
al. (1997) to convert luminosity to absolute visual magnitude.

There are a number of other groups that have produced publicly
available, extensive databases of tracks and isochrones that may be
used as the grid for grid modelling efforts. These models often use
different model parameters and physics input. We use two such
available sets of models, one described by Dotter et al.~(2008) and
the other by Marigo et al.~(2008).

The Dotter et al. grid is a collection of stellar evolution tracks and
isochrones which were computed with the Dartmouth Stellar Evolution
code (DSEP; Dotter et al. 2007). The input physics used by DSEP is
similar to YREC. The modelling parameters are somewhat different. They
assume a mixing length parameter of $\alpha=1.938$.  The extent of
core-overshoot is assumed to be a function of mass, with overshoot
ramping up from 0.05$H_p$ to 0.2$H_p$. We only use a subset of their
models, specifically the ones with [$\alpha$/Fe]=0 and  [Fe/H] of $-$1.0,
$-$0.5, 0.0, 0.2, 0.3, and 0.5. The initial helium abundance of the
models is $Y = 0.245 + 1.54 Z$ and they also assume the solar
metallicity of Grevesse \& Sauval (1998). The models range in mass
from 0.1 to 5 $M_{\odot}$ in increments of 0.05 $M_{\odot}$ in the
range 0.1 to 1.8 $M_{\odot}$, increments of 0.1 $M_{\odot}$ for masses
between 1.8 to 3.0 $M_{\odot}$ and increments 0.2 $M_{\odot}$ for
higher masses.  The models were downloaded from the DSEP
web-page.\footnote{http://stellar.dartmouth.edu/\~{}models/index.html}

The Marigo et al.  grid consists of models with the Padova stellar
evolution code (Marigo et al. 2008; Girardi et al. 2000).  Padova use
the OPAL high temperature opacities (Rogers \& Iglesias 1992; Iglesias
\& Rogers 1993) complemented in the low temperature regime with the
tables of Alexander \& Ferguson (1994). They assume a mixing length
parameter of $\alpha=1.68$ and have both envelope and core convective
overshoot and both are functions of mass. They define their solar
metallicity to be $Z=0.019$ following Grevesse \& Noels (1993) and
$Y=0.273$, whilst also adopting a somewhat complicated helium
enrichment model.  The models we use in our grid were downloaded from
the Padova CMD web
page.\footnote{http://stev.oapd.inaf.it/cgi-bin/cmd}

\section{Results}
\label{sec:res}

We used about 7300 simulated stars, drawn from the YREC grid described
above, to test the grid method. 
The simulated stars are a subset of models drawn at random from the model grid of 820,000 stars, 
in such a way as to obtain a homogeneous sampling in mass, age and metallicity.
Determining the parameters of these
test stars with the YREC grid will reveal systematic errors inherent
in the grid method, while using these with the other grids described
below will reveal the model-dependence of the results, including
effects of using different values of the mixing length parameter. We
estimate the radius, mass, \logg\, and age of each simulated star
twice, once with error-free data to test the systematic errors due to
the method and once with random noise added to the inputs to determine
the errors that we can expect, and in particular, how the random and
systematic errors interact.  For this work we assume errors of 2.5\%
in \dnu, 5\% in \numax, 100K in \teff\ and 0.1 dex in [Fe/H].
 Radius and mass estimates obtained from error-free data using the
direct method are expected to be exact (solving
two equations for two unknowns). The situation is less
clear for the grid methods since the relationship between
mass, radius and temperature for stellar models is more complex
as well as non-linear.

\subsection{Radius}

\begin{figure}
\epsscale{0.9} \plotone{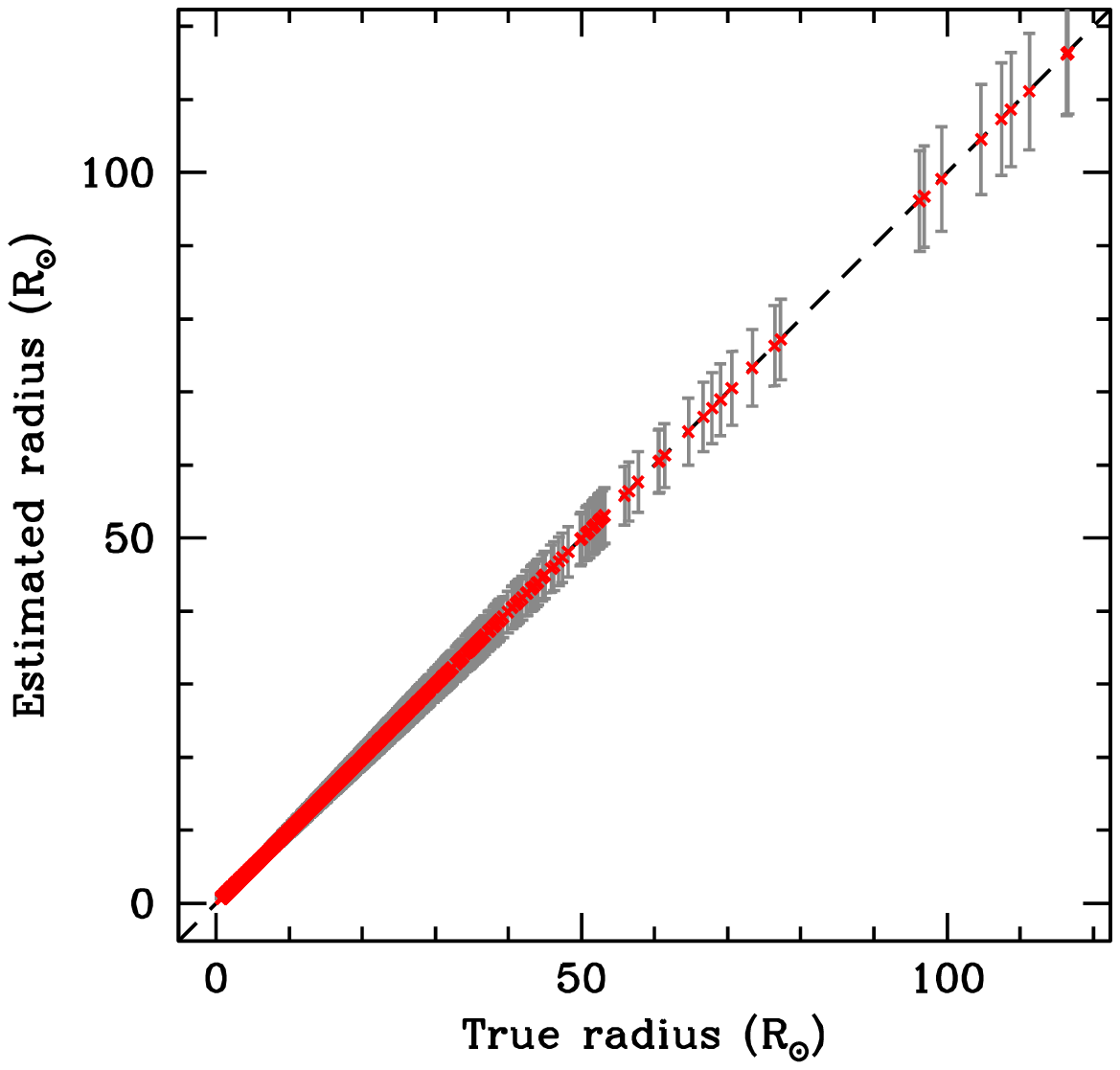} \caption{Radius of
simulated stars obtained $\Delta\nu$, $\nu_{\rm max}$ and $T_{\rm
eff}$ using Equations~(\ref{eq:delnu}) and (\ref{eq:numax}). No
errors were added to the data. The gray errorbars show what the
errors would be assuming a 2.5\% error in $\Delta\nu$, 5\% error in
$\nu_{\rm max}$ and 100K error in $T_{\rm eff}$.}
\label{fig:ne_rad_dir}
\end{figure}

\begin{figure}
\epsscale{0.55} \plotone{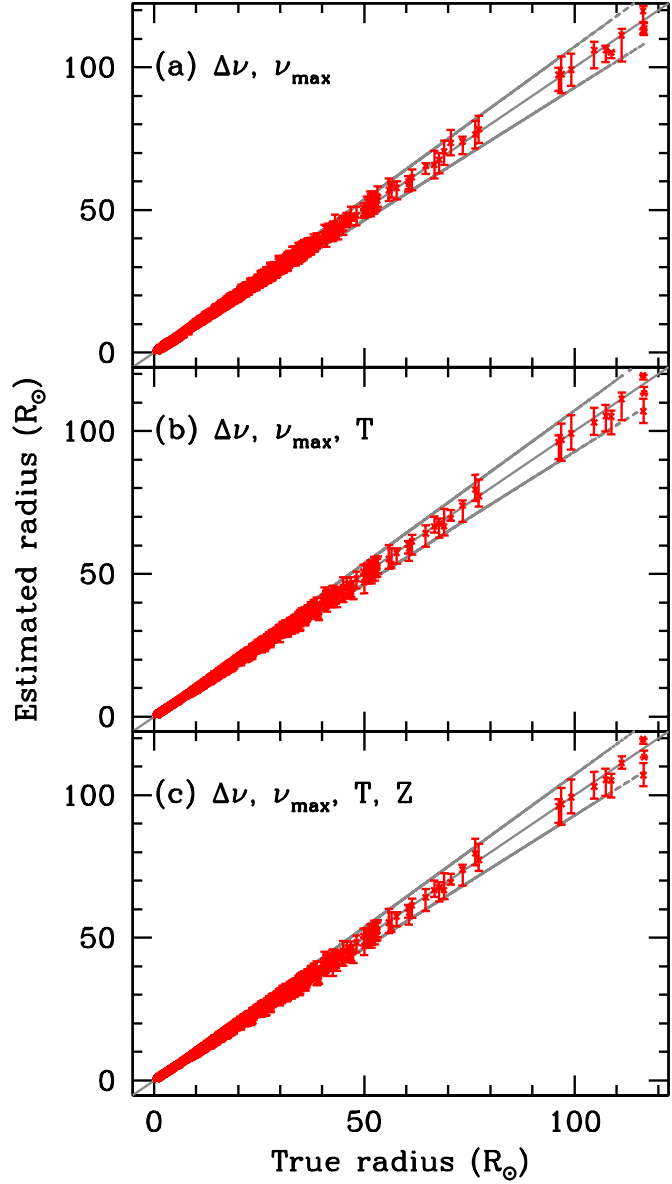} \caption{Radius of
simulated stars obtained using the grid method using various
combinations of seismic and non-seismic data. No errors were added
to the data. The points with errorbars show the estimated radii and
the errorbars we would have expected for the errors adopted. The
middle gray line is where the points would be if the method worked
perfectly, while the two other show the $1\sigma$ errors using the
``direct'' method. Note that the grid method results in somewhat
smaller errorbars, but that even for data with no errors, we do not
always estimate the radius correctly. } \label{fig:ne_rad_y_grid}
\end{figure}

\begin{figure}
\epsscale{0.9} \plotone{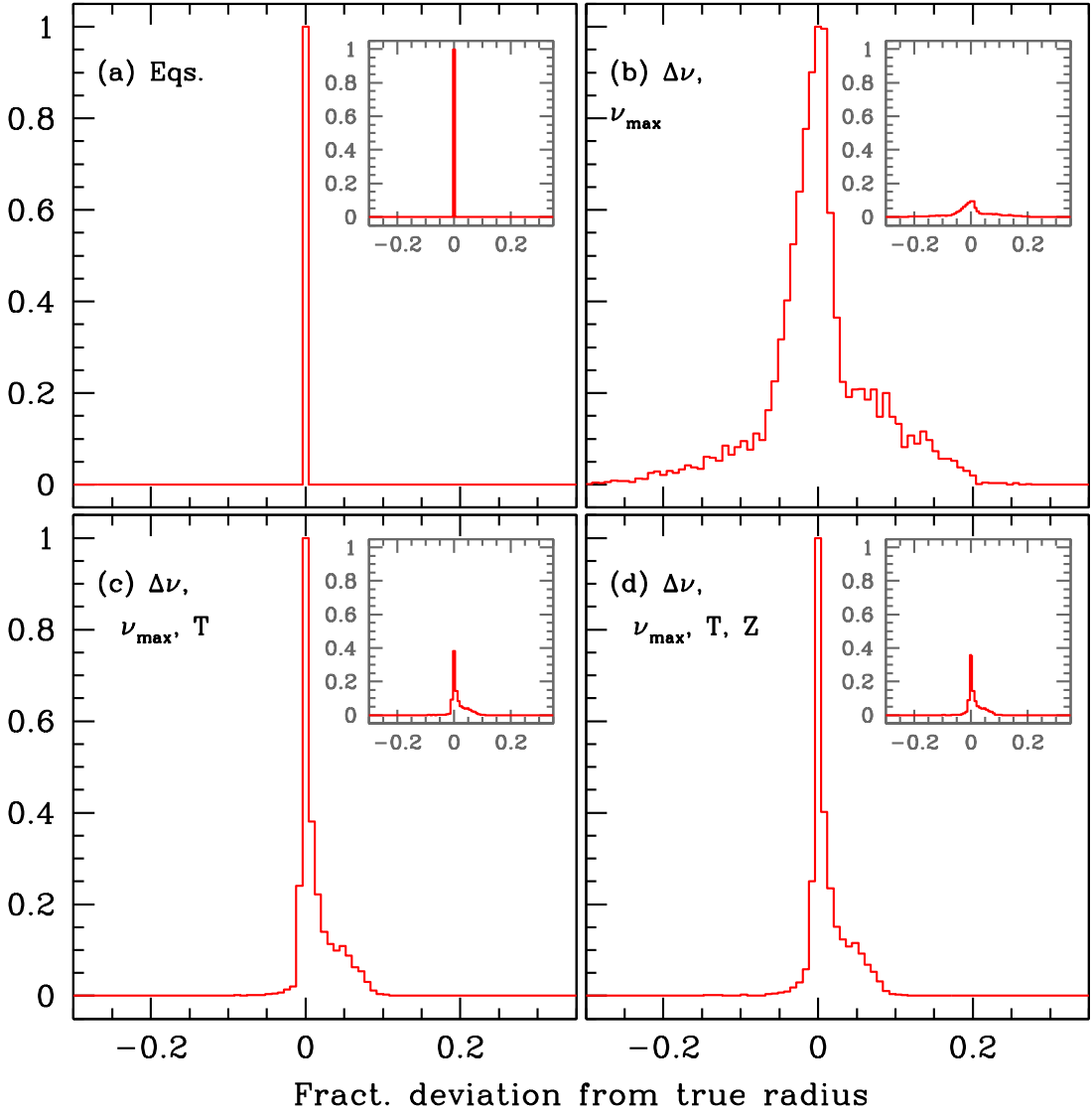} \caption{Histograms
showing the fractional deviation between the true radii and the
radii estimated using different methods to show the accuracy of the
methods. These results are for error-free data.
All histograms have been normalized to unity at the maximum to
facilitate easy comparison of their widths. Panel (a) is the result
of the direct method, which as can be seen, gives a perfect results.
Panels (b), (c) and (d) are results for the grid method using
different combinations of data as mentioned in the figure legends.
 The inset in each panel
shows the distributions normalized to unit area.
}
\label{fig:ne_rad_yrec_hist}
\end{figure}

We show the result of directly estimating radius using
Equations~\ref{eq:delnu} and \ref{eq:numax} in
Figure~\ref{fig:ne_rad_dir}. These are results for error-free data,
however, we also show the error-bars that would result if we had used
data with the errors we have adopted. The radius estimates obtained
with the grid method, again for error-free data, are shown in
Figure~\ref{fig:ne_rad_y_grid}. These are results obtained with the
YREC grid. We show results of using different combinations of
inputs. It should be noted that although the input data have no
errors, the adopted errors were used to define the volume in the grid
within which we calculated the likelihood function. As can be seen,
there is some systematic error in the results, in particular at large
radii where the estimated radius differs slightly from the true
radius. 
 It can be seen that the uncertainties in the results obtained
by the grid method will  be lower than those obtained by the direct method
when inputs that have errors are used.
The systematic error is quantified
in Figure~\ref{fig:ne_rad_yrec_hist} where we have plotted a
normalized histogram of the deviation of the estimated radius from the
true radius for each of the four cases shown in
Figures~\ref{fig:ne_rad_dir} and \ref{fig:ne_rad_y_grid}.

Note that we have plotted histograms normalized to unity at maximum.
While normalizing to unit area emphasizes differences between the distributions
 (wider distributions have smaller amplitude), it is difficult to estimate
visually, and compare, the full-width at half maximum (FWHM) of the distributions.
 The FWHM of the distributions tells us whether or not the
error-distribution is acceptable. Normalizing the distributions to have a maximum 
value of unity makes the differences between the distributions less obvious. However, it makes
determining (and comparing) the FWHM of different distributions much easier. 
Since it is often conventional to plot histograms normalized to unit area, in
each figure we show an inset with the conventional normalization.

As can be seen from Figure~\ref{fig:ne_rad_yrec_hist}, the direct method has no systematic error.
The systematic errors in the grid method depend on the combination of
inputs used, and it is evident that it is important to know \teff\ in
addition to the seismic quantities. The addition of metallicity to the
inputs does not appear to make much of a difference.

\begin{figure}
\epsscale{0.9} \plotone{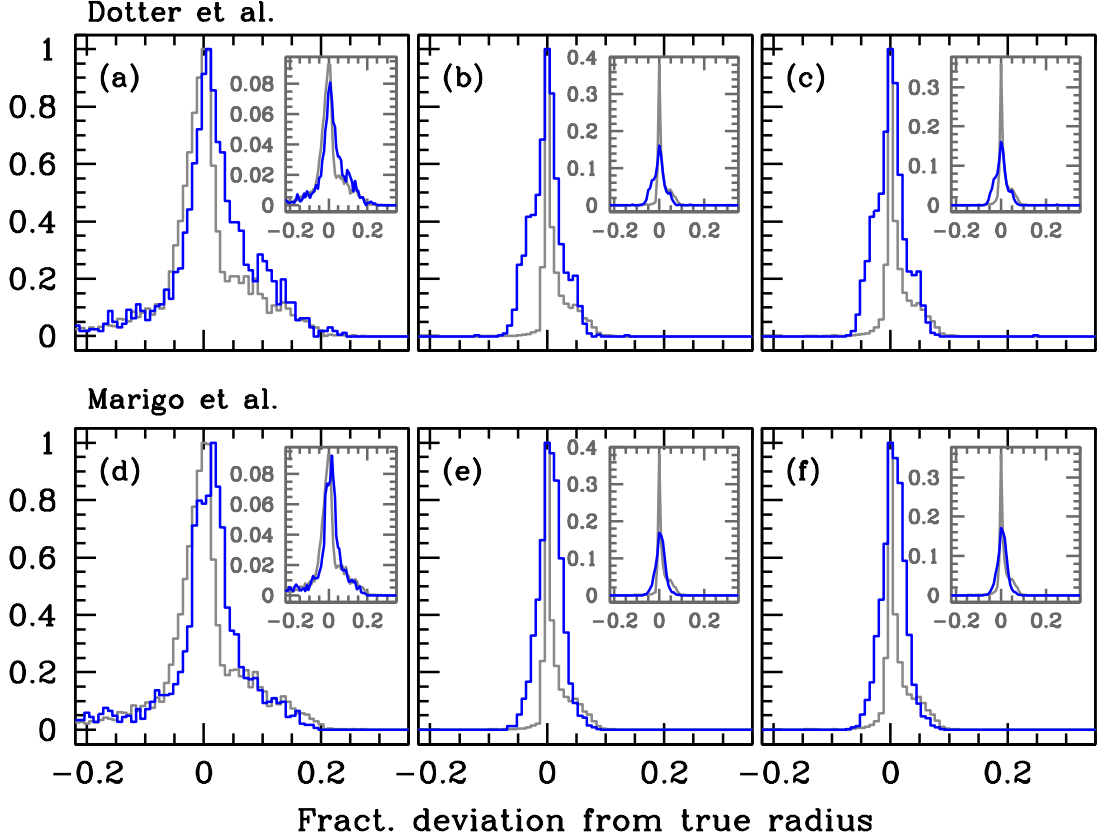} \caption{The effect
of using a grid constructed with somewhat different physics than the
simulated stars (which were all YREC models). As in
Figure~\ref{fig:ne_rad_yrec_hist} we show histograms showing the
fractional deviation between the true and estimated radii. In blue
we show results obtained using the grid of Dotter et al. models
(upper row) and the Marigo et al. models (lower row). The histograms
in gray show the corresponding results for the YREC grid.  Results are
for error-free data and show that the effects of known uncertainties
in stellar models can lead to systematic errors in the results
obtained using the grid method. Panels (a) and (d) are results
obtained using \dnu\ and \numax; panels (b) and (e) are for \dnu, \numax, 
and \teff; panels (c) and (f) are results for \dnu, \numax, \teff and [Fe/H].
 The inset in each panel
shows the distributions normalized to unit area.
} \label{fig:ne_rad_other_hist}
\end{figure}

The results described above were obtained for YREC-based stars while
using the YREC grid, and hence, a grid with the same physics.  In order
to test possible systematic errors induced by uncertainties in physics
as well as uncertainties in modelling parameters such as the mixing
length parameter, we repeated the grid modelling exercise for YREC
stars with Dotter et al. and Marigo et al. grids. Again we used
error-free data. As mentioned earlier, the Dotter et al. and Marigo et
al. models differ considerably from YREC models. 
The histogram of
errors for the radii of YREC stars, as obtained using the two other
grids, are shown in Figure~\ref{fig:ne_rad_other_hist}. As can be
seen, the errors in the results are large; however, once \teff\ is
known, the half-width at half-maximum (HWHM) is only about 2.5\%,
though the largest errors can be about 10\%. This error is of the same
order as that obtained for the YREC grid when we did not use
\teff. These results are consistent with our earlier results in Basu
et al.~(2010) for radius determinations with models with different
mixing lengths, however, while we had earlier considered only six
cases, two each at the main-sequence, turn-off and sub-giant phases,
we have now considered a few thousand of cases with models spanning
all evolutionary stages from the zero-age main-sequence to the
red-clump.

\begin{figure}
\epsscale{0.9} \plotone{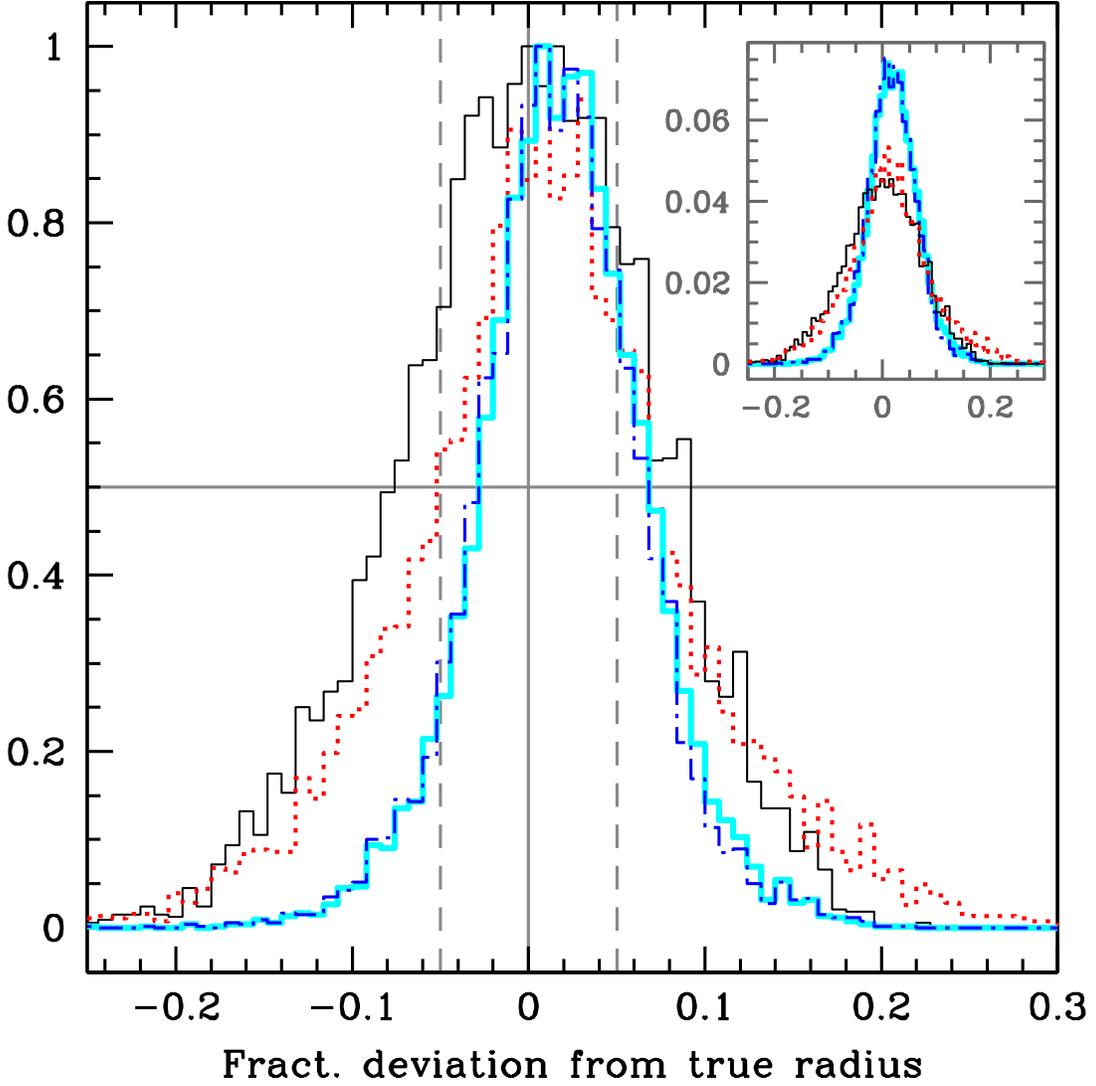} \caption{
Histograms showing the fractional deviation between the true radii and
radii obtained from different methods when errors were 
added to the data. These histograms show the precision of the method.
The results were obtained using the YREC grid.  The thin black solid
line shows the result of using the direct method. The other lines
are for the grid method using different data combinations: the red
dotted line shows the result of
using only $\Delta\nu$ and $\nu_{\rm max}$; the thick cyan line 
for $\Delta\nu$, $\nu_{\rm max}$ and $T_{\rm eff}$; 
and the blue dot-dashed line  for  $\Delta\nu$, $\nu_{\rm max}$, 
$T_{\rm eff}$ and $Z$.
The solid gray vertical line indicates zero deviation, while the
gray dashed lines indicate errors of $\pm5$\%. The grey horizontal line
marks the half-maximum of the distributions. 
 The inset shows the distributions normalized to unit area.
}
\label{fig:e_rad_yrec_hist}
\end{figure}

Our results give confidence that the grid method works well for
estimating the radius, at least where error-free data are
concerned. The question, of course, is what happens if data with errors
are used. The results for data with our fiducial errors obtained using
the YREC grid are shown in Figure~\ref{fig:e_rad_yrec_hist}. Results
for about 7300 stars are shown in the histograms. One can see that the
direct method gives the worst results. Even if we only use the two
seismic parameters \dnu\ and \numax, the grid method gives better
results than the direct method. When \teff\ and metallicity are also
used the HWHM is about 5\%. In contrast, the HWHM is about 10\% when
the direct method is used. While using only \dnu\ and \numax\ in the
grid method gives a low HWHM (about 7\%), the distribution has
substantial tails showing that for individual stars we could get large
errors in radius if we only use the two seismic inputs. However, for
the nominal errors that we have adopted, the errors we obtain are
rarely greater than 20\%.

\begin{figure}
\epsscale{0.8} \plotone{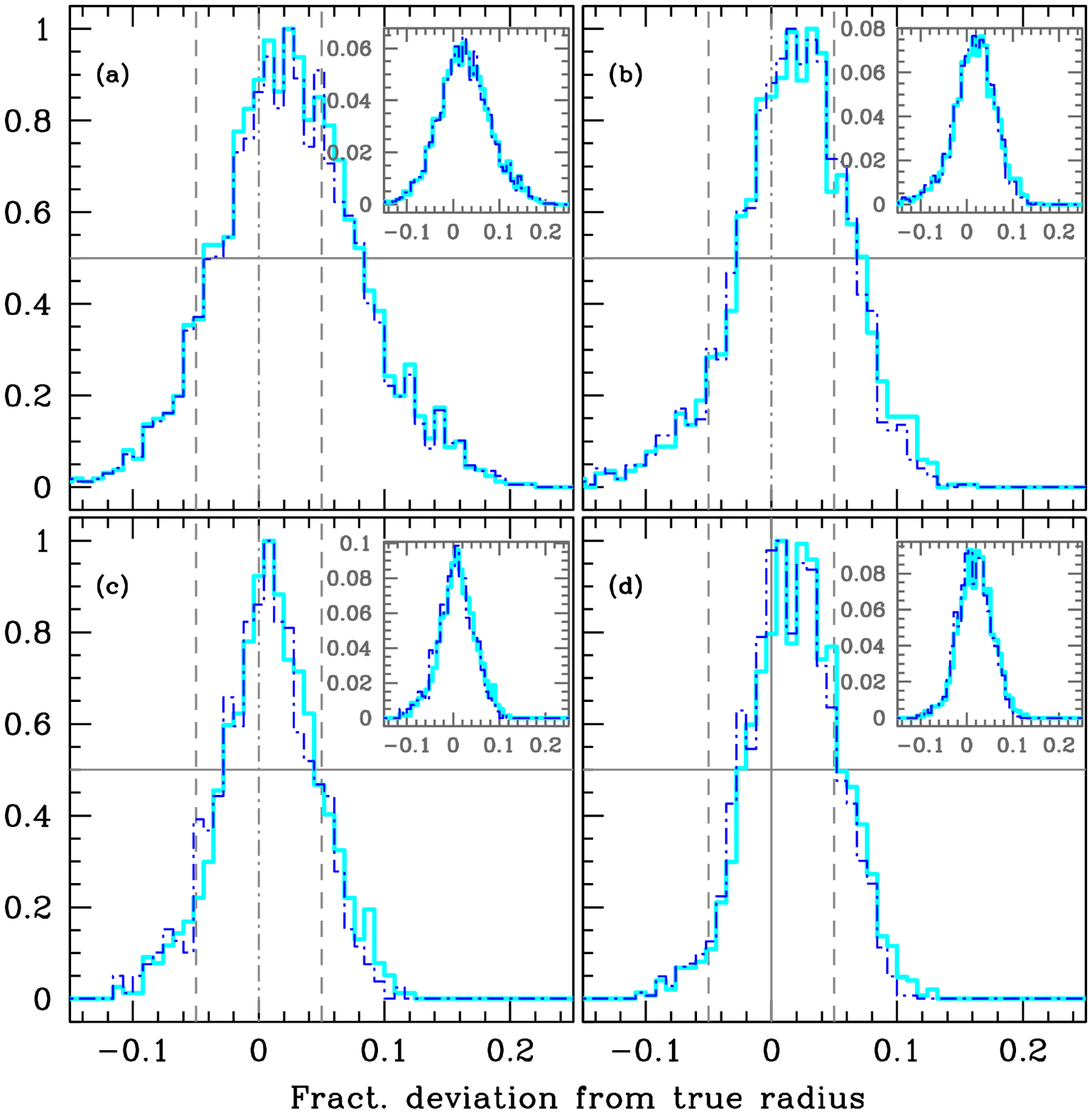}
\caption{The same as Figure~\ref{fig:e_rad_yrec_hist}, but for
stars in selected $\Delta\nu$ ranges. We show results for
stars with $\Delta\nu \le 20\mu$Hz (panel a), $ 20 < \Delta\nu \le
75\mu$Hz (panel b), $75 < \Delta\nu \le 100\mu$Hz (panel c), and $
\Delta\nu >  100\mu$Hz (panel d). We only show the grid results
using the $(\Delta\nu, \nu_{\rm max}, T_{\rm eff})$ combination
(thick cyan line) and the $(\Delta\nu, \nu_{\rm max}, T_{\rm eff},
Z)$ combination (dot-dashed blue line). Only results obtained with
YREC data are shown.
 The inset in each panel shows the
distributions normalized to unit area.
}
\label{fig:e_rad_yrec_sel_hist}
\end{figure}

How well we can estimate the radius of a star appears to depend on its
evolutionary state. In Figure~\ref{fig:e_rad_yrec_sel_hist} we show
the errors in the radius when we split our sample into four groups by
\dnu. Stars with $\Delta\nu\leq 20\mu$Hz, i.e., giants, give the worst
estimates --- this result is consistent what we reported in Basu et
al.~(2010). The best estimates are obtained for stars with
$75\leq\Delta\nu<100\mu$Hz.

All distributions of errors obtained for the grid method are
asymmetric about the zero point, which reflects the complex non-linear
relationship between radius and the observed stellar parameters. With
the errors adopted for this work, the figures tell us that we can
expect errors in radius estimates caused by errors in the
observational inputs to dominate over the systematic errors of the
grid method. However, that will not be the case if we can decrease
errors in \dnu\ and \numax\ by a much larger amount, an unlikely case
for {\it Kepler} survey stars. Random errors in temperature have a
much smaller effect and are relatively unimportant. And since it is
unlikely that we can get metallicity errors to be lower than the
adopted value of 0.1\,dex, we will probably always be random-error
dominated.

\begin{figure}
\epsscale{0.6} \plotone{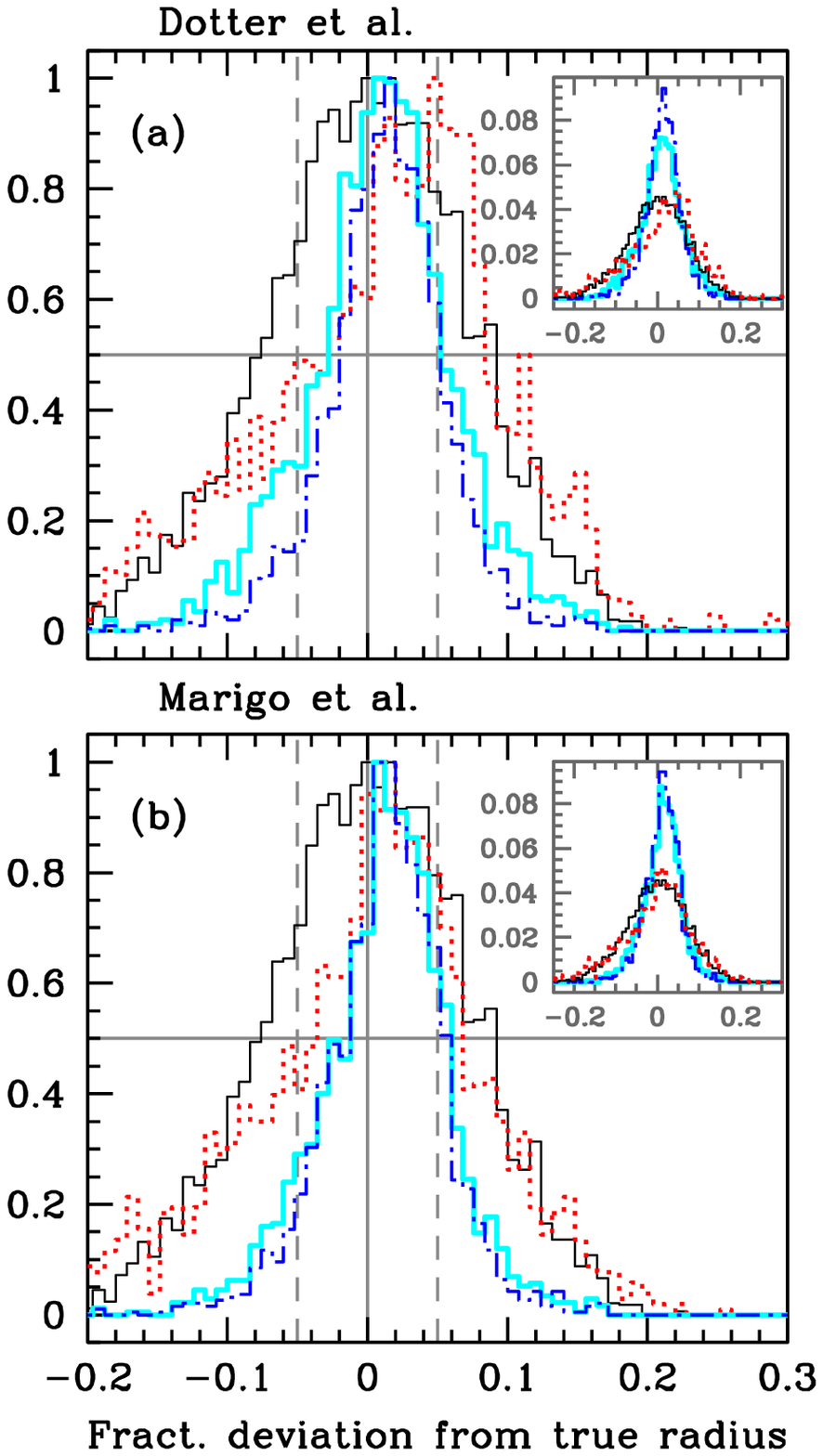} \caption{Same as
Figure~\ref{fig:e_rad_yrec_hist} but for results obtained using the
grid of Dotter et al. models (panel a) and Marigo et al. models
(panel b). The line types are also the same as those in
Figure~\ref{fig:e_rad_yrec_hist}. } 
\label{fig:e_rad_other_hist}
\end{figure}

Using dissimilar grids does change the situation somewhat, in
particular, there is a large systematic shift in the results if only
the two seismic parameters \dnu\ and \numax\ are used as shown in
Figure~\ref{fig:e_rad_other_hist}. The systematic shift becomes
insignificant when \teff\ is used and reduces further when metallicity
is also used. The HWHM of the error-distribution, however, is
comparable to what is obtained with the YREC grid, i.e., about
5\%.

Given that 1D stellar models can never simulate real stars,
we try out the grid method on solar data using all three grids of
models. We used the solar large spacing obtained from solar
$\ell=0$ frequencies measured by the Birmingham Solar-Oscillations
Network (Chaplin et al. 1996). In particular we use the mode set
BiSON-1 described in Basu et al. (2009) to determine an the average
large spacing calculated between 2.47\,mHz and 3.82\,mHz for use as a
seismic input parameter.  The interval was chosen to be roughly ten
large spacings centred around the frequency of maximum power. This
choice is prompted by what we can expect from Kepler in the Survey
Phase. The errors used were the same as that adopted in this paper.
The results are tabulated in Table~\ref{tab:rad}, and as can be
seen, we can determine the solar radius quite well. The errors
are smallest when we use the combination (\dnu, \numax, \teff, and 
$Z$).
Since using dissimilar grids simulates some of the uncertainties we
may face when we get real data, and since we can reproduce the solar radius
well using \dnu\ and \numax\ for the Sun using all three grids, we can be confident
about our error estimates.

\begin{table}
\caption{Solar radius (in units of $R_\odot$) obtained with BiSON data using different grids and 
input combinations}
\begin{center}
\begin{tabular}{lccc}
\hline
\noalign{\smallskip}
Grid  & \multispan3{\hfill Input combinations\hfill}\\
 & (\dnu, \numax) & (\dnu, \numax, \teff) & (\dnu, \numax, \teff, $Z$) \\
\noalign{\smallskip}
\hline
\noalign{\smallskip}
Dotter et al. & $0.978^{+0.020}_{-0.018}$& $1.008^{+0.051}_{-0.062}$& $1.002^{+0.042}_{-0.054} $\\
Marigo et al. & $1.016^{+0.021}_{-0.016}$& $1.000^{+0.036}_{-0.051}$& $1.001^{+0.035}_{-0.049} $\\
YREC &          $1.016^{+0.021}_{-0.020}$& $0.992^{+0.029}_{-0.050}$& $1.001^{+0.065}_{-0.053} $\\
\noalign{\smallskip}
\hline
\end{tabular}
\end{center}
\label{tab:rad}
\end{table}

\subsection{Mass}
\label{subsec:mass}

We have repeated the tests described above for the case of
determining masses of stars.

\begin{figure}
\epsscale{0.9} \plotone{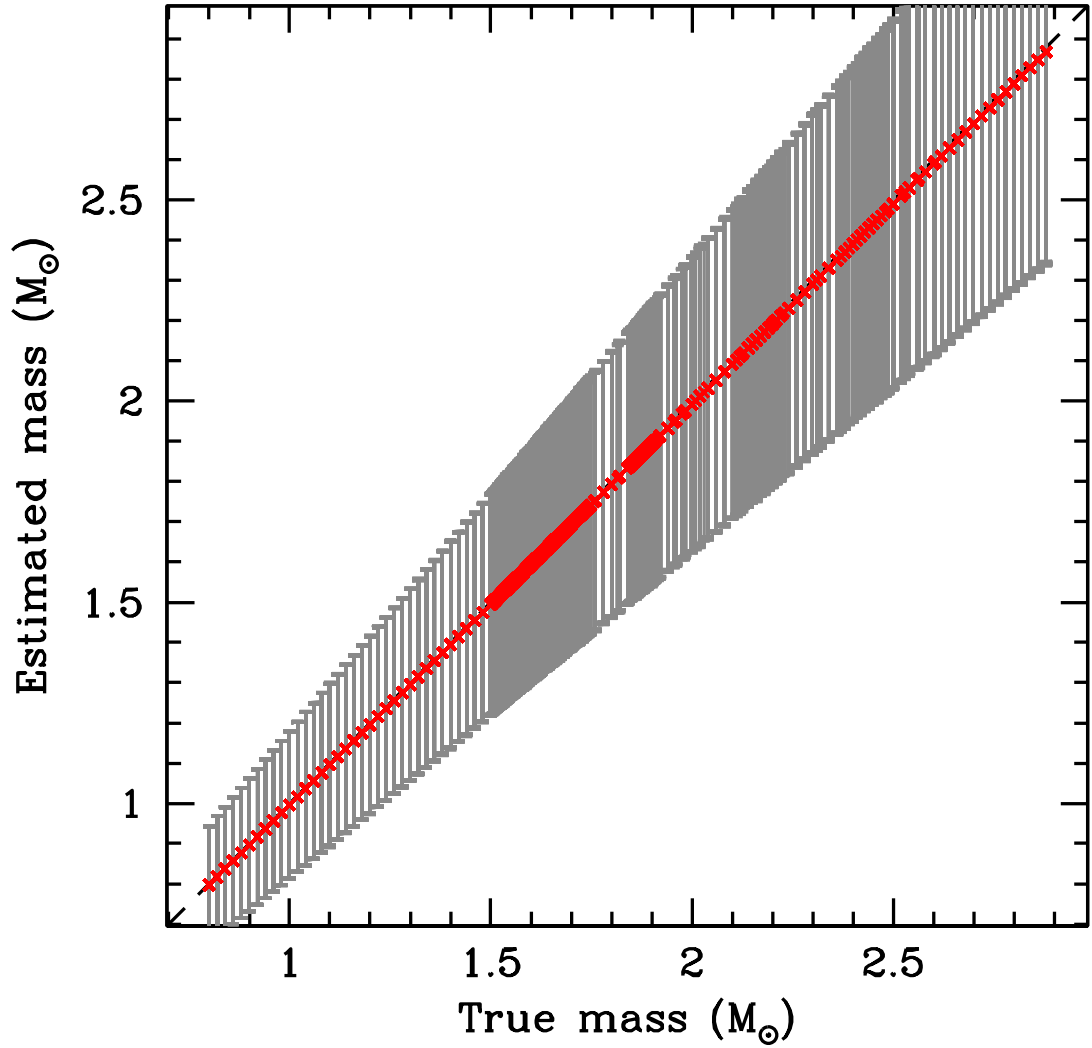} 
\caption{Mass of
simulated stars obtained from $\Delta\nu$, $\nu_{\rm max}$ and $T_{\rm
eff}$ using Equations~(\ref{eq:delnu}) and (\ref{eq:numax}). No
errors were added to the data. The gray errorbars show what the
errors would be for our fiducial errors.
 } 
\label{fig:ne_mass_dir}
\end{figure}

\begin{figure}
\epsscale{0.6} \plotone{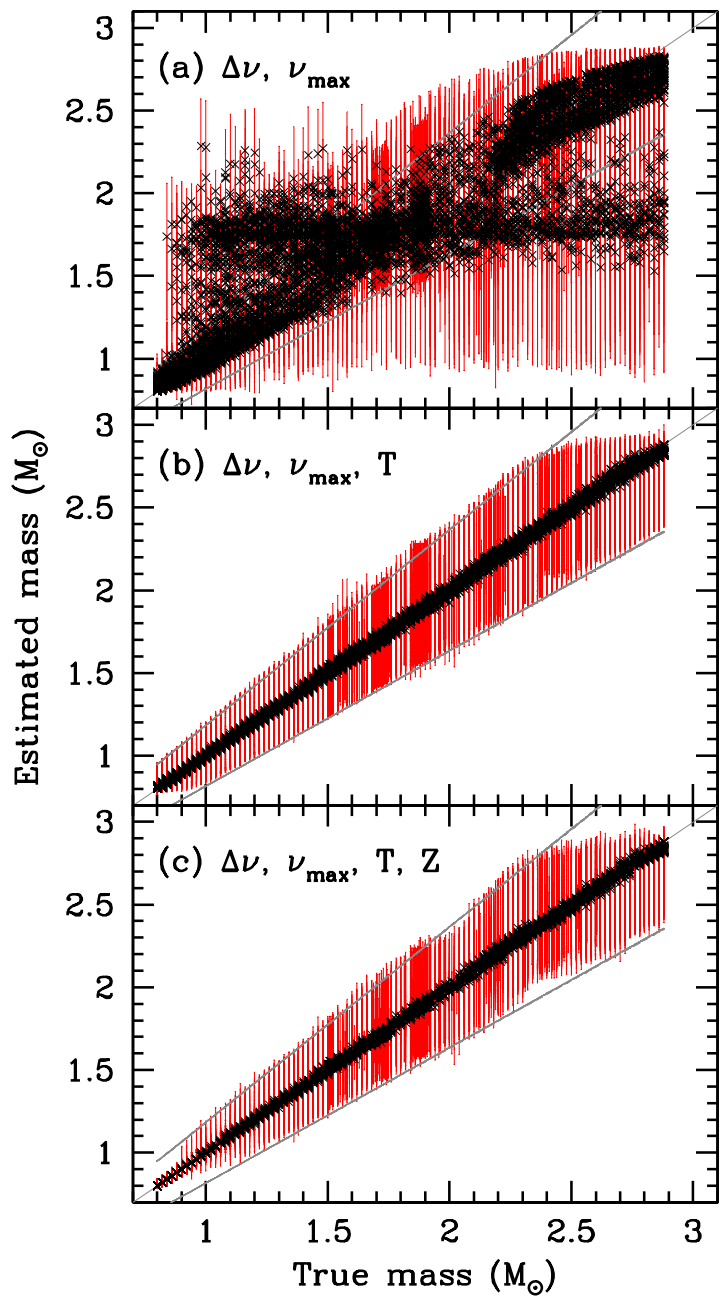} \caption{Mass of
simulated stars obtained using the grid method. No errors were added
to the data. The black  points with errorbars show the estimated radii and
the errorbars we would have expected for the errors adopted. The
middle gray line is where the points would be if the method worked
perfectly, while the two other show the $1\sigma$ errors using the
``direct'' method. All results were obtained with the YREC grid.}
\label{fig:ne_mass_y_grid}
\end{figure}

\begin{figure}
\epsscale{0.9} \plotone{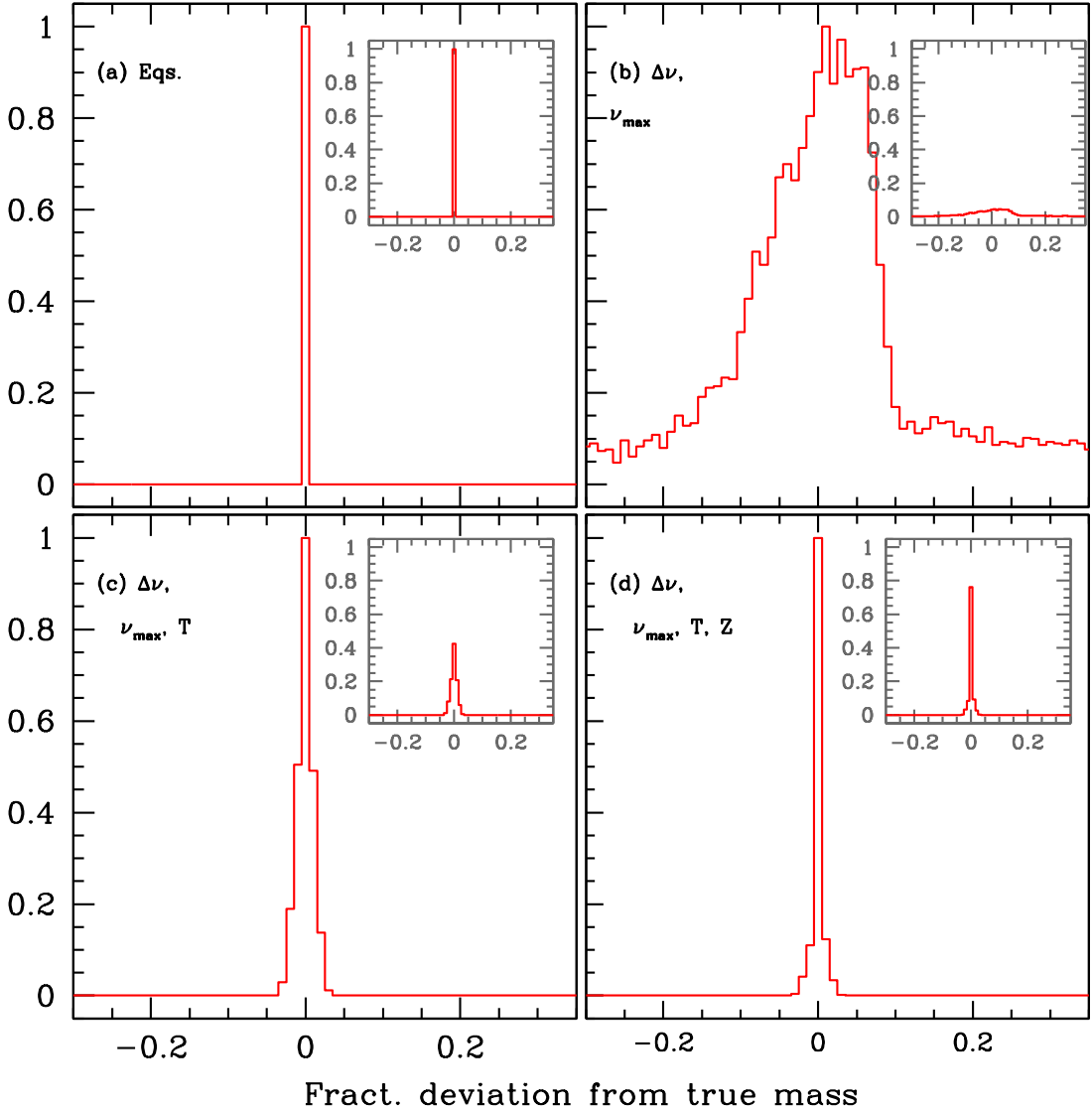} \caption{Histograms
showing the fractional deviation between the true masses and the masses 
estimated using different methods to show the accuracy of the
methods. These results were obtained using the YREC grid with error-free data.
All histograms have been normalized to unity at the maximum to
facilitate easy comparison of their widths. Panel (a) is the result
of the direct method, which as can be seen, gives a perfect results.
Panels (b), (c) and (d) are results for the grid method using
different combinations of data as mentioned in the figure legends.
 The inset in each panel show the distributions normalized to
unit area.}
\label{fig:ne_mass_yrec_hist}
\end{figure}

\begin{figure}
\epsscale{0.9} \plotone{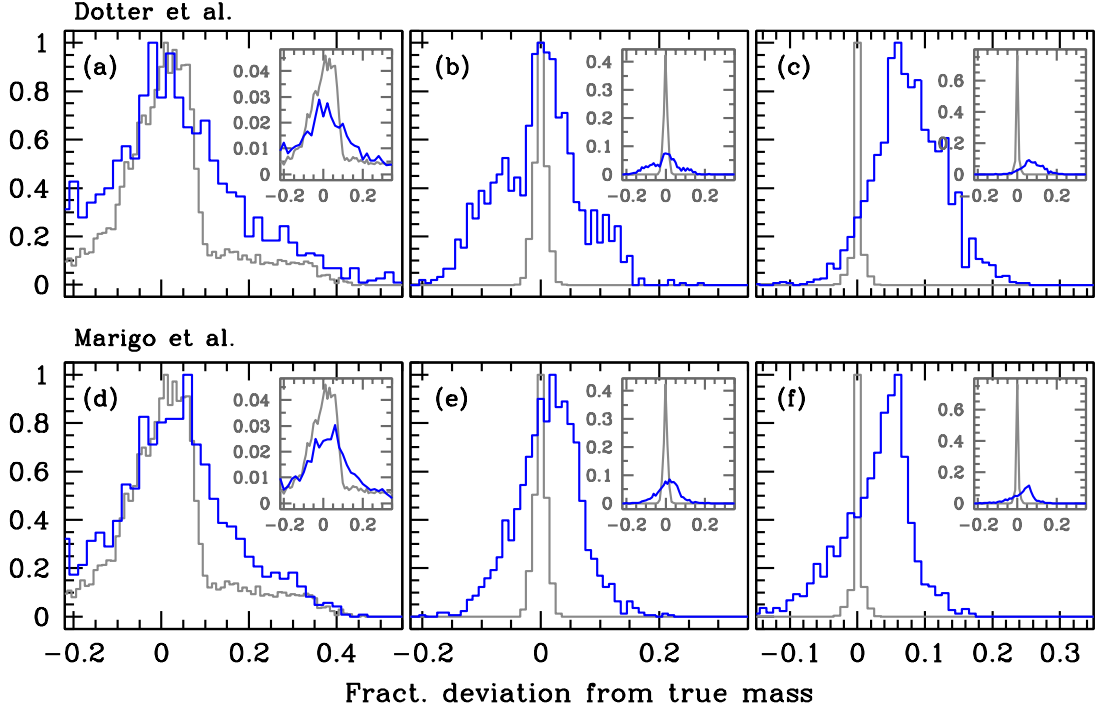} \caption{The effect
of using a grid constructed with somewhat different physics than the
simulated stars (which were all YREC models). In blue we show the
histograms of fractional deviation between the true and estimated masses
obtained  using the grid of Dotter et al. models
(upper row) and the Marigo et al. models (lower row). The histograms
in gray show the corresponding results for the YREC grid. Results are
for error-free data. Panels (a) and (d) are results
obtained using \dnu and \numax; panels (b) and (e) are for \dnu, \numax,
and \teff; panels (c) and (f) are results for \dnu, \numax, \teff and [Fe/H].
 The insets show the
distributions normalized to unit area.
}
\label{fig:ne_mass_other_hist}
\end{figure}

The direct method, as is obvious from Equations~(\ref{eq:delnu}) and
(\ref{eq:numax}), gives much larger error in mass, and this can be seen
in Figure~\ref{fig:ne_mass_dir} where we show the results for the
error-free case. The expected errors reduce (but become asymmetric)
when the grid method with YREC models is used. The grid results in the
error-free case are shown in Figure~\ref{fig:ne_mass_y_grid}, and we
find that unless we know \teff, we cannot use the grid method.  Also,
as far as mass is concerned, the grid method is most useful for the
lowest-mass stars, and also stars with somewhat higher masses,
particularly if metallicity is known. The error-distributions for the
different cases are shown in Figure~\ref{fig:ne_mass_yrec_hist}. It is
very clear that using only \dnu\ and \numax\ does not work. 
 Again the question is what will happen in the real case. Since we do not have independent
mass and seismic measurements of stars other than the Sun, we mimic this case by
using a grid of models that are different from the proxy stars.
The results are not very promising, as can be
seen from Figure~\ref{fig:ne_mass_other_hist} and it appears that it
is essential that we have good metallicity estimates.

\begin{figure}
\epsscale{0.9} \plotone{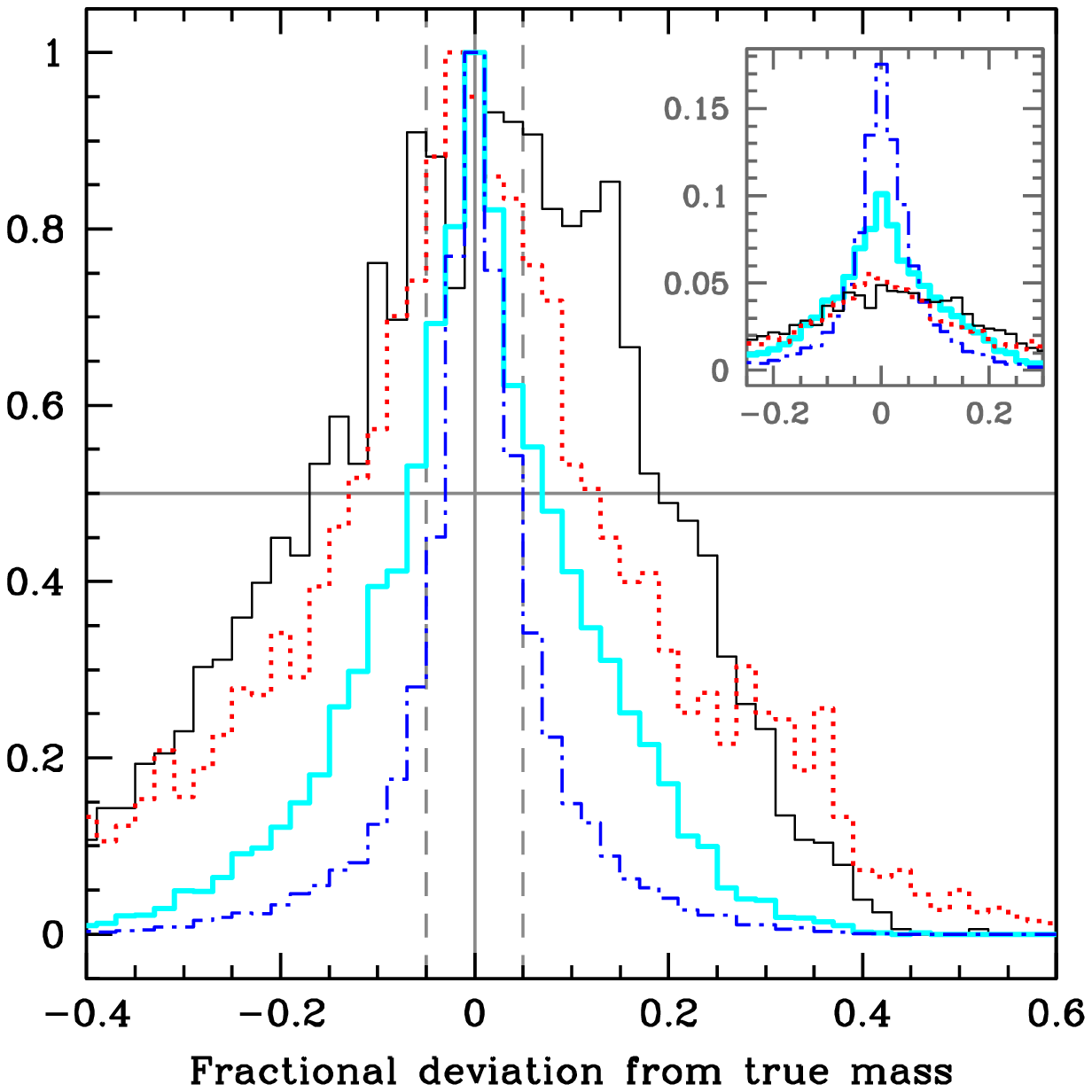} \caption{
Histograms showing the fractional deviation between the true masses and
masses obtained from different methods when errors were 
added to the data.  The thin black solid
line shows the result of using the direct method. The other lines
are for the grid method using the YREC grid and different data combinations: 
the red dotted line shows the result of
using only $\Delta\nu$ and $\nu_{\rm max}$; the thick cyan line
for $\Delta\nu$, $\nu_{\rm max}$ and $T_{\rm eff}$;  
and the blue dot-dashed line  for  $\Delta\nu$, $\nu_{\rm max}$,
$T_{\rm eff}$ and $Z$.
The solid gray vertical line indicates zero deviation, while the
gray dashed lines indicate errors of $\pm5$\%. The grey horizontal line
marks the half-maximum of the distributions. Note that the
distributions are wider than the corresponding distributions for
radii (Fig.~\ref{fig:e_rad_yrec_hist}).
 The inset shows the distributions normalized to unit area.
}
\label{fig:e_mass_yrec_hist}
\end{figure}

\begin{figure}
\epsscale{0.6} \plotone{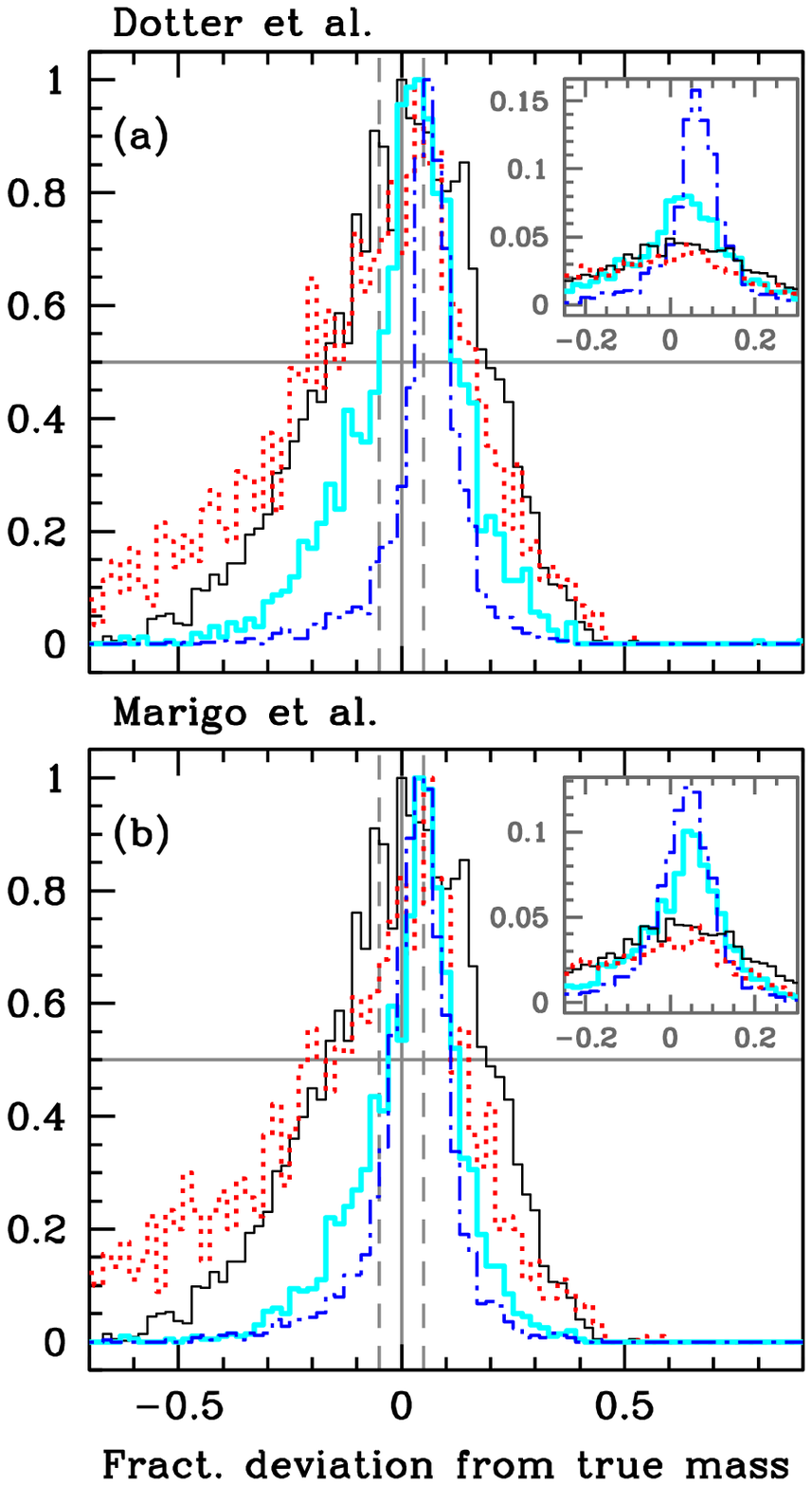} \caption{The same as
Figure~\ref{fig:e_mass_yrec_hist} but for results obtained using the
grid of Dotter et al. models (panel a) and Marigo et al. models
(panel b). The line types are also the same as those in
Figure~\ref{fig:e_mass_yrec_hist}. 
}
\label{fig:e_mass_other_hist}
\end{figure}

Despite the seemingly discouraging results, it still appears that in
the  more realistic case, i.e., when the input data have errors, grid modelling
will give better results. The error-distribution for the direct
method, and the grid method for three different inputs, is shown in
Figure~\ref{fig:e_mass_yrec_hist} for the YREC grid and in
Figure~\ref{fig:e_mass_other_hist} for the Dotter et al. and Marigo et
al. grids. It is clear that as long as we know the effective
temperature, final errors in the grid modelling will be much better
than those in the direct determination case. The results are
substantially improved when metallicity is known, with a HWHM of
around 5\%.   Solar mass
estimates using the different grid are listed in Table~\ref{tab:mass},
and we do indeed get the best results if we explicitly use the knowledge of
metallicity.

\begin{table}
\caption{Solar mass (in units of $M_\odot$) obtained with BiSON data using different grids and 
input combinations}
\begin{center}
\begin{tabular}{lccc}
\hline
\noalign{\smallskip}
Grid  & \multispan3{\hfill Input combinations\hfill}\\
 & (\dnu, \numax) & (\dnu, \numax, \teff) & (\dnu, \numax, \teff, $Z$) \\
\noalign{\smallskip}
\hline
\noalign{\smallskip}
Dotter et al. & $1.066^{+0.116}_{-0.069}$& $1.025^{+0.125}_{-0.225}$& $1.000^{+0.050}_{-0.050}$\\
Marigo et al. & $1.032^{+0.066}_{-0.074}$& $1.004^{+0.094}_{-0.130}$& $1.012^{+0.066}_{-0.062}$\\
YREC &          $1.067^{+0.057}_{-0.171}$& $1.002^{+0.183}_{-0.152}$& $1.027^{+0.052}_{-0.060}$\\
\noalign{\smallskip}
\hline
\end{tabular}
\end{center}
\label{tab:mass}
\end{table}

\begin{figure}
\epsscale{0.8} \plotone{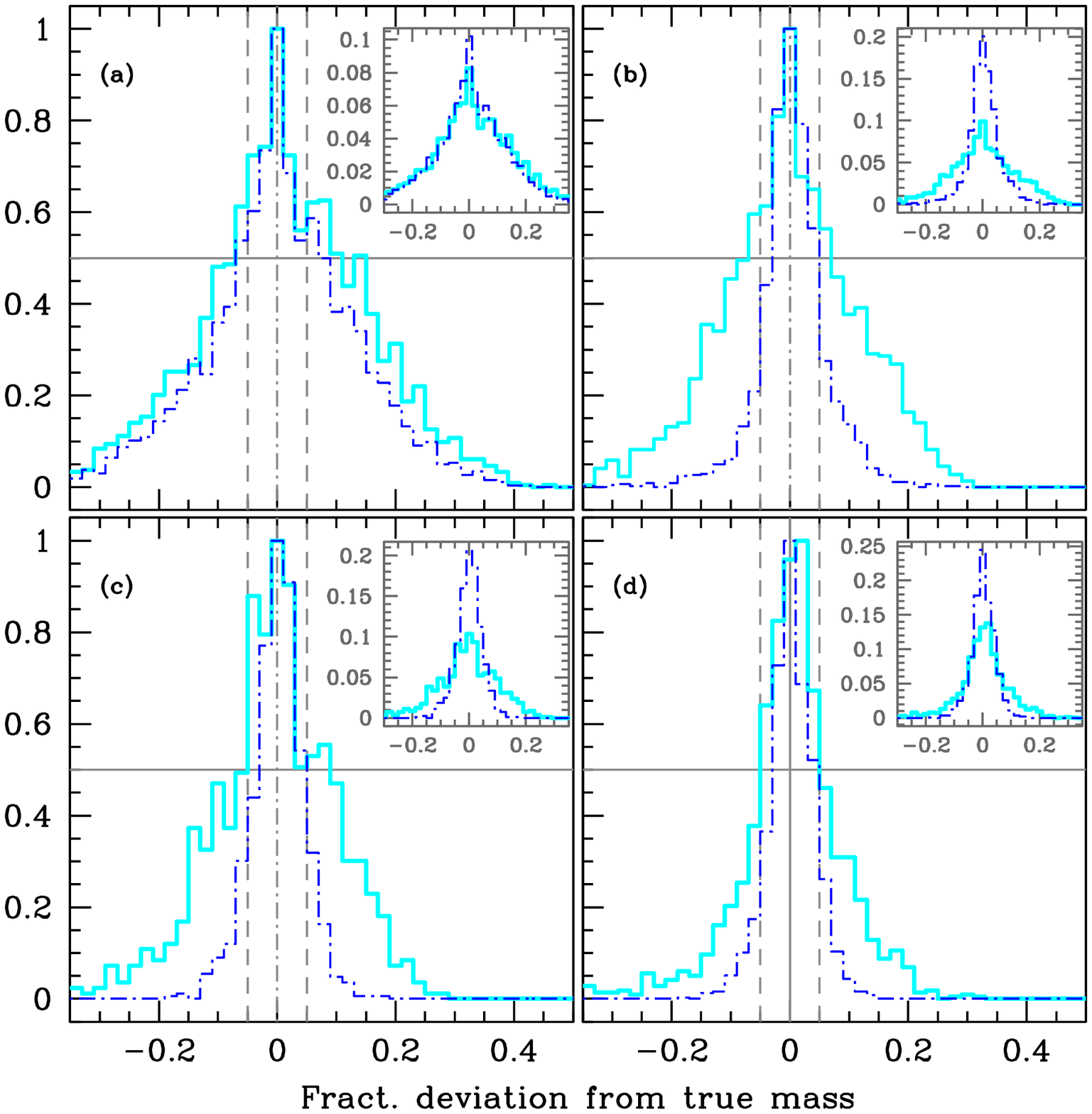} \caption{
The same as Figure~\ref{fig:e_mass_yrec_hist}, but for
stars in selected $\Delta\nu$ ranges. We show results for
stars with $\Delta\nu \le 20\mu$Hz (panel a), $ 20 < \Delta\nu \le
75\mu$Hz (panel b), $75 < \Delta\nu \le 100\mu$Hz (panel c), and $
\Delta\nu >  100\mu$Hz (panel d). We only show the grid results
using the $(\Delta\nu, \nu_{\rm max}, T_{\rm eff})$ combination
(thick cyan line) and the $(\Delta\nu, \nu_{\rm max}, T_{\rm eff},
Z)$ combination (dot-dashed blue line).
 The inset in each panel shows the distributions normalized to unit area.
}
\label{fig:e_mass_yrec_sel_hist}
\end{figure}

As in the case of radius, it is easier to find masses of stars that
are either on or close to the main-sequence than it is for red-giants.
Error-distributions of mass estimates obtained in different ranges of
\dnu\ are shown in Figure~\ref{fig:e_mass_yrec_sel_hist}. Only results
obtained with the YREC grid are shown.

\subsection{Error correlations and \logg}
\label{subsec:err}

\begin{figure}
\epsscale{0.9} \plotone{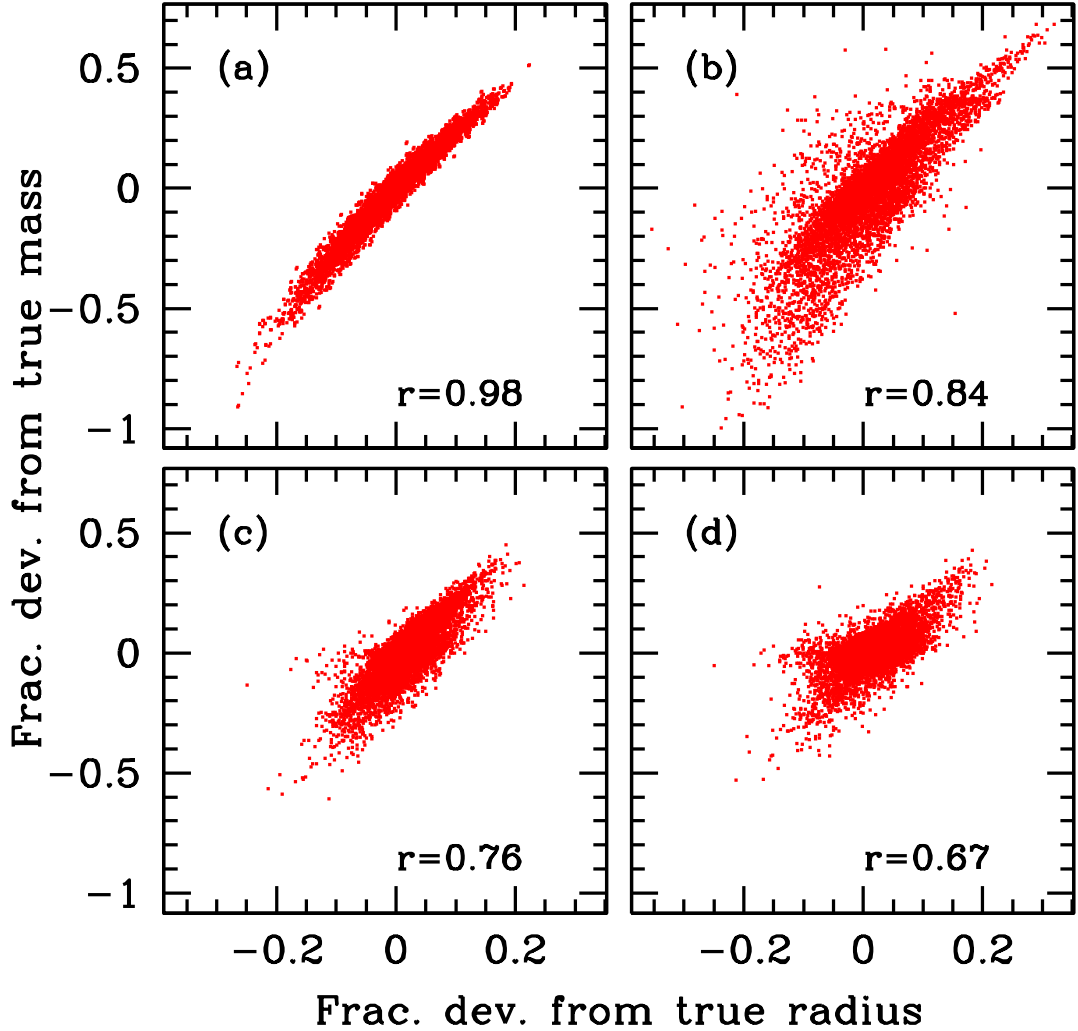} \caption{The
correlation between errors in estimated mass and radius as given by
different methods. The linear correlation coefficient in each case
is mentioned in the figure. Panel (a) is the result of the direct
method, which, as can be seen, gives perfect results. Panels (b),
(c) and (d) are results for the grid method using different
combinations of data as mentioned in the figure legends: panel (b)
for $(\Delta\nu, \nu_{\rm max})$, panel (c) for $(\Delta\nu,
\nu_{\rm max}, T_{\rm eff})$ and panel (d) for $(\Delta\nu, \nu_{\rm
max}, T_{\rm eff}, Z)$. Only results obtained with YREC grid
are shown.} \label{fig:e_corr_yrec_hist}
\end{figure}

We expect errors in mass and radius to be correlated. In the direct
method the two quantities are determined from the same two equations,
and in the grid method they are determined from the same population of
models. While we can expect the errors to be almost completely
correlated for the direct method, the expected correlation is less
obvious for the grid method. To determine the extent of the
correlations we have plotted the fractional deviation from true mass
against the fractional deviation from true radius for the nearly 7300
simulated stars. The results are shown in
Figure~\ref{fig:e_corr_yrec_hist}. The errors are
positively correlated. The linear correlation coefficient is noted in
each panel and as can be seen, the correlation is substantial even in
the grid modelling case. Thus, we need to keep in mind that if we
underestimate the radius, we will also underestimate the mass of the
star in question.

\begin{figure}
\epsscale{0.9} \plotone{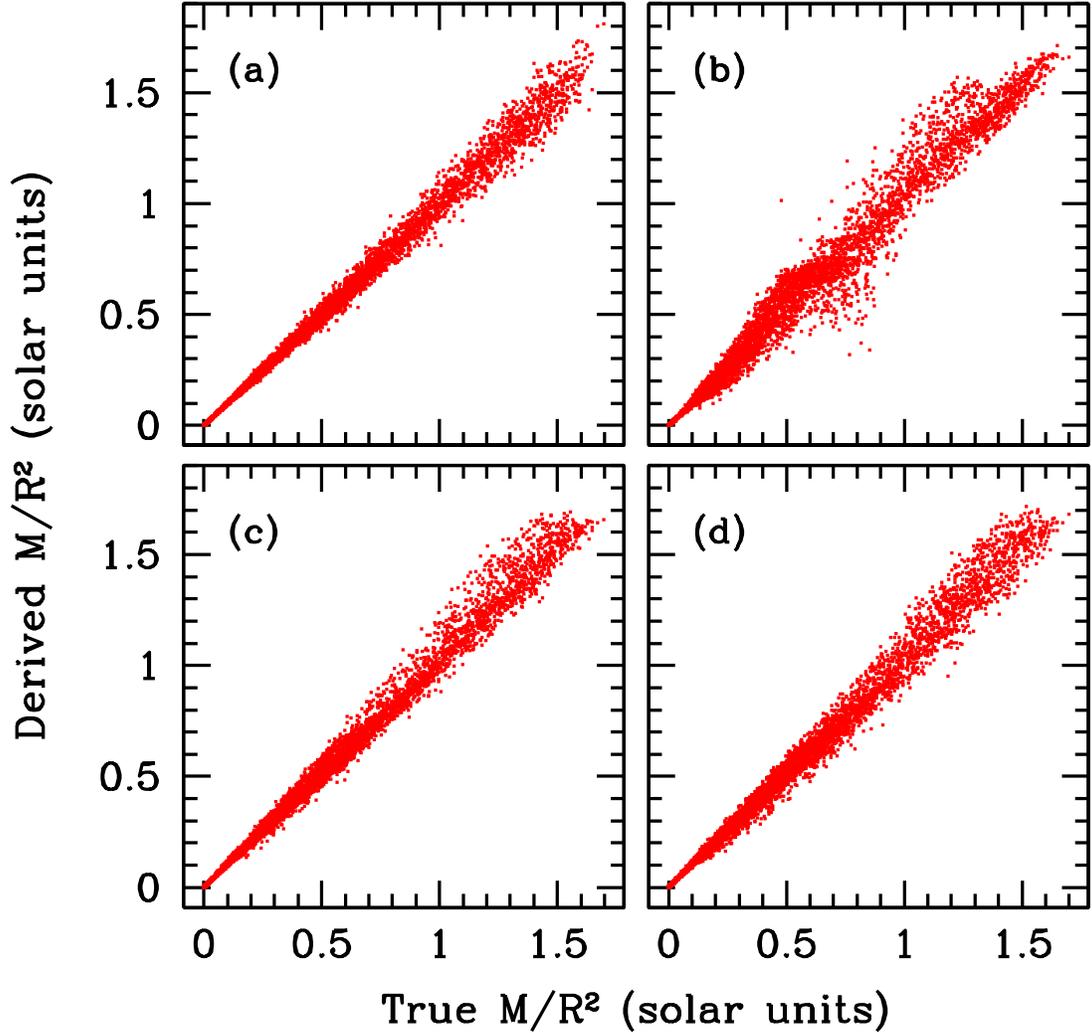} \caption{ The ratio
$M/R^2$ derived from estimated values of mass and radius and plotted as
a function of the true value. The results indicate that we should be
able to estimate \logg\ accurately. Panel (a) is the result of the
direct method, which, as can be seen, gives  perfect results. Panels
(b), (c) and (d) are results for the grid method using different
combinations of data as mentioned in the figure legends: panel (b)
for $(\Delta\nu, \nu_{\rm max})$, panel (c) for $(\Delta\nu,
\nu_{\rm max}, T_{\rm eff})$ and panel (d) for $(\Delta\nu, \nu_{\rm
max}, T_{\rm eff}, Z)$.} \label{fig:e_mr2}
\end{figure}

The positive correlation between the mass and radius pointed us to the
fact that errors in functions involving the ratio of $M$ and $R$ may
be smaller.  From Figure~\ref{fig:e_corr_yrec_hist} it can be seen
that when the deviation of mass is plotted against the deviation of
radius, we do not get a completely straight line but a somewhat curved
relation and since $M/R$ is not an observed physical quantity, we
investigate $M/R^2$, i.e., $g$, instead. In Figure~\ref{fig:e_mr2} we
plot the value of $M/R^2$ derived from $M$ and $R$ estimated
separately with the true value of $M/R^2$ and find that regardless of
the method, or of the combination of inputs to the grid method, we get a tight
straight line. We, therefore, believe that we should be able to estimate
\logg\ of stars very well.

\begin{figure}
\epsscale{0.9} \plotone{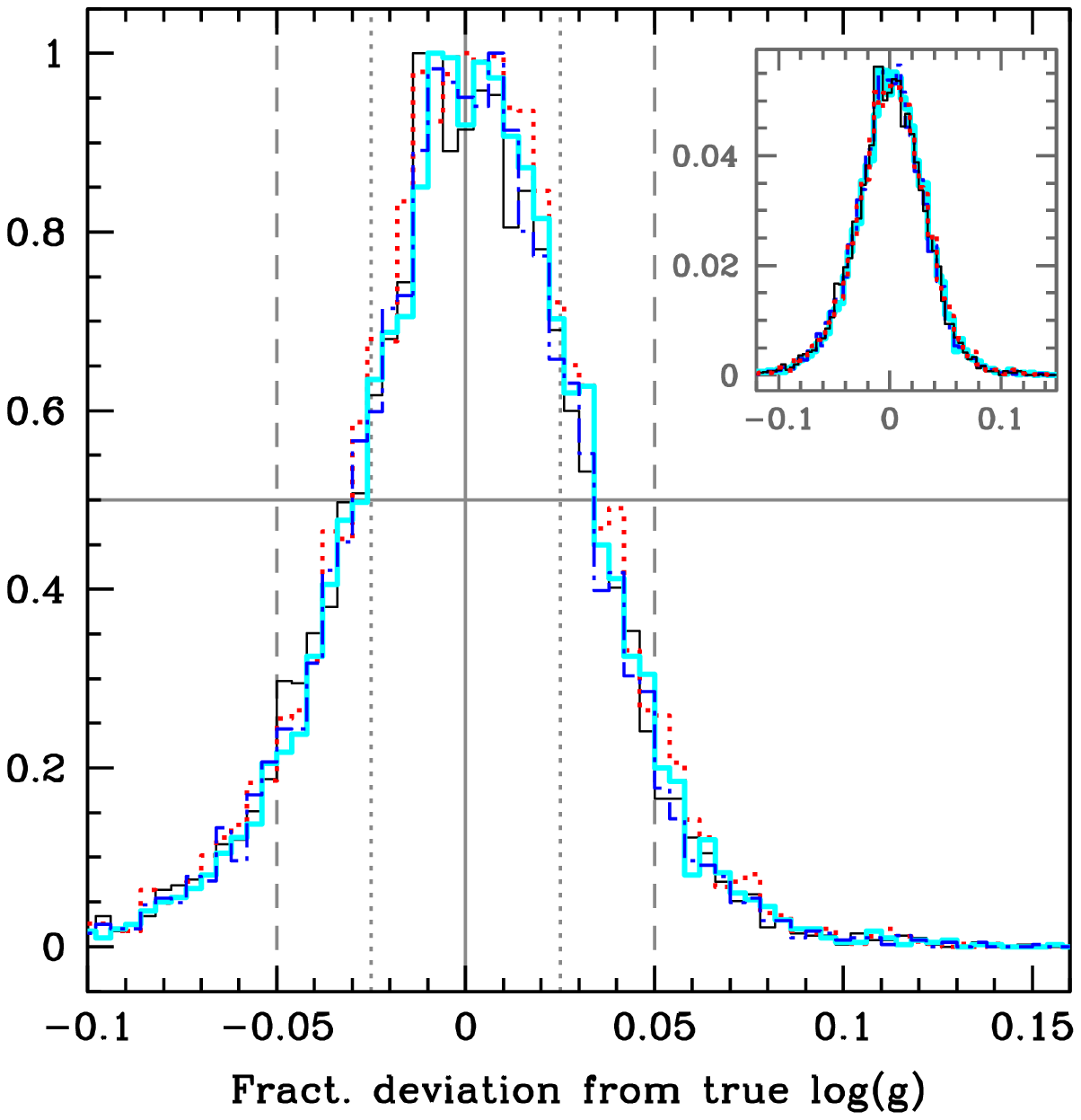} \caption{
Histograms showing the fractional deviation between true \logg\ and
\logg\ obtained from different methods using the YREC grid when errors were
added to the data. The thin black solid
line shows the result of using the direct method. The other lines
are for the grid method using different data combinations: the red
dotted line shows the result of
using only $\Delta\nu$ and $\nu_{\rm max}$; the thick cyan line
for $\Delta\nu$, $\nu_{\rm max}$ and $T_{\rm eff}$;
and the blue dot-dashed line  for  $\Delta\nu$, $\nu_{\rm max}$,
$T_{\rm eff}$ and $Z$.
The solid gray vertical line indicates zero deviation, while the
gray dashed lines indicate errors of $\pm5$\%, the  dotted gray 
lines represent $\pm 2.5$\% error and the grey horizontal line
marks the half-maximum of the distributions.
 The inset shows the
distributions normalized to unit area.}
\label{fig:e_g_yrec_hist}
\end{figure}

\begin{figure}
\epsscale{0.6} \plotone{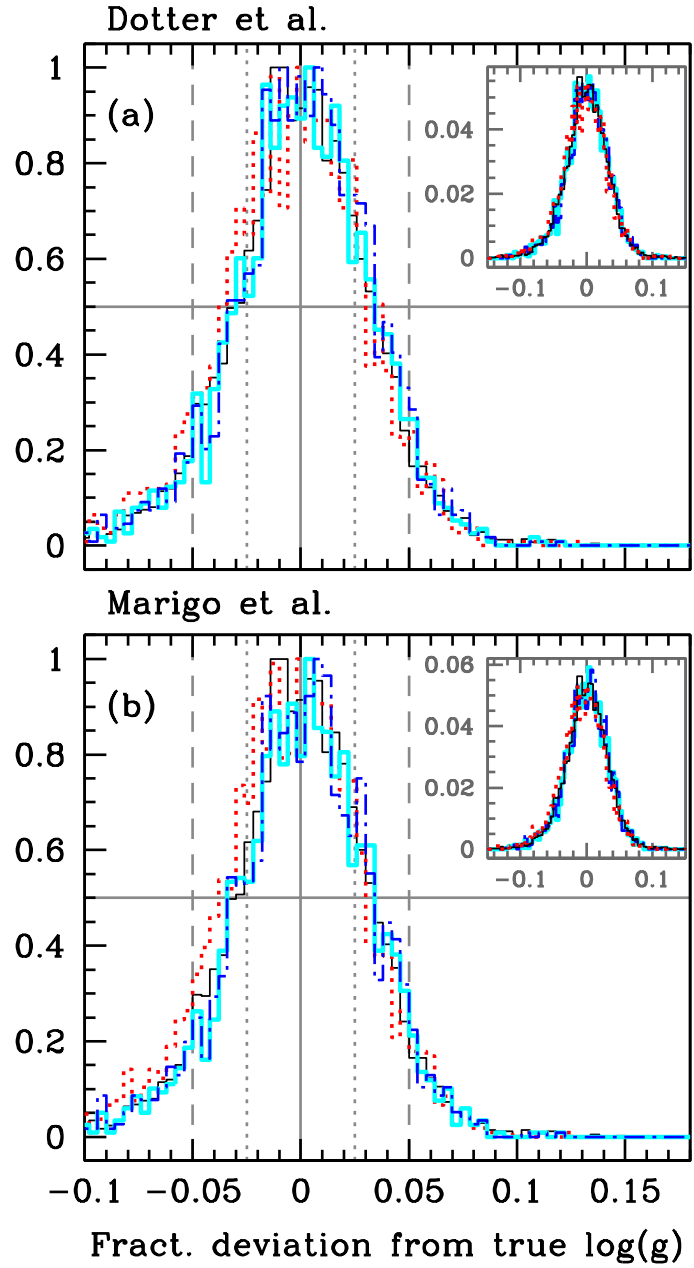} \caption{
The effect of dissimilar models on \logg\ estimates. The line types are
the same as in Figures~\ref{fig:e_rad_other_hist} and \ref{fig:e_mass_other_hist}. 
 The additional dotted gray lines represent $\pm 2.5$\%
error. } \label{fig:e_g_other_hist}
\end{figure}

The error-histograms for \logg\ determined by different methods, using
the YREC grid, are shown in Figure~\ref{fig:e_g_yrec_hist} and with
the Marigo et al. and Dotter et al. grids are shown in
Figure~\ref{fig:e_g_other_hist}.  We only show the case where the
input quantities had errors added.  As can be seen, all methods give
almost equally good results and when similar models are used the HWHM
is only 2.5\%. The error is slightly larger (HWHM about 3.5\%) when
dissimilar models are used.

 As with radius and mass, we have used BiSON data to estimate \logg\ for the Sun. 
The results are listed in Table~\ref{tab:logg}. As expected from the discussion above,
we get good results no matter which input combination we use.

\begin{table}
\caption{Sun's \logg\ obtained with BiSON data using different grids and 
input combinations. Note that the accepted value of solar \logg\ is 4.438.}
\begin{center}
\begin{tabular}{lccc}
\hline
\noalign{\smallskip}
Grid  & \multispan3{\hfill Input combinations\hfill}\\
 & (\dnu, \numax) & (\dnu, \numax, \teff) & (\dnu, \numax, \teff, $Z$) \\
\noalign{\smallskip}
\hline
\noalign{\smallskip}
Dotter et al. &$4.447^{+0.016}_{-0.014}$& $4.438^{+0.018}_{-0.021}$ & $4.437^{+0.013}_{-0.011}$\\
Marigo et al. &$4.442^{+0.011}_{-0.012}$& $4.436^{+0.016}_{-0.016}$ & $4.438^{+0.014}_{-0.015}$\\
YREC &         $4.444^{+0.012}_{-0.016}$& $4.434^{+0.016}_{-0.016}$ & $4.443^{+0.014}_{-0.013}$\\
\noalign{\smallskip}
\hline
\end{tabular}
\end{center}
\label{tab:logg}
\end{table}

\begin{figure}
\epsscale{0.8} \plotone{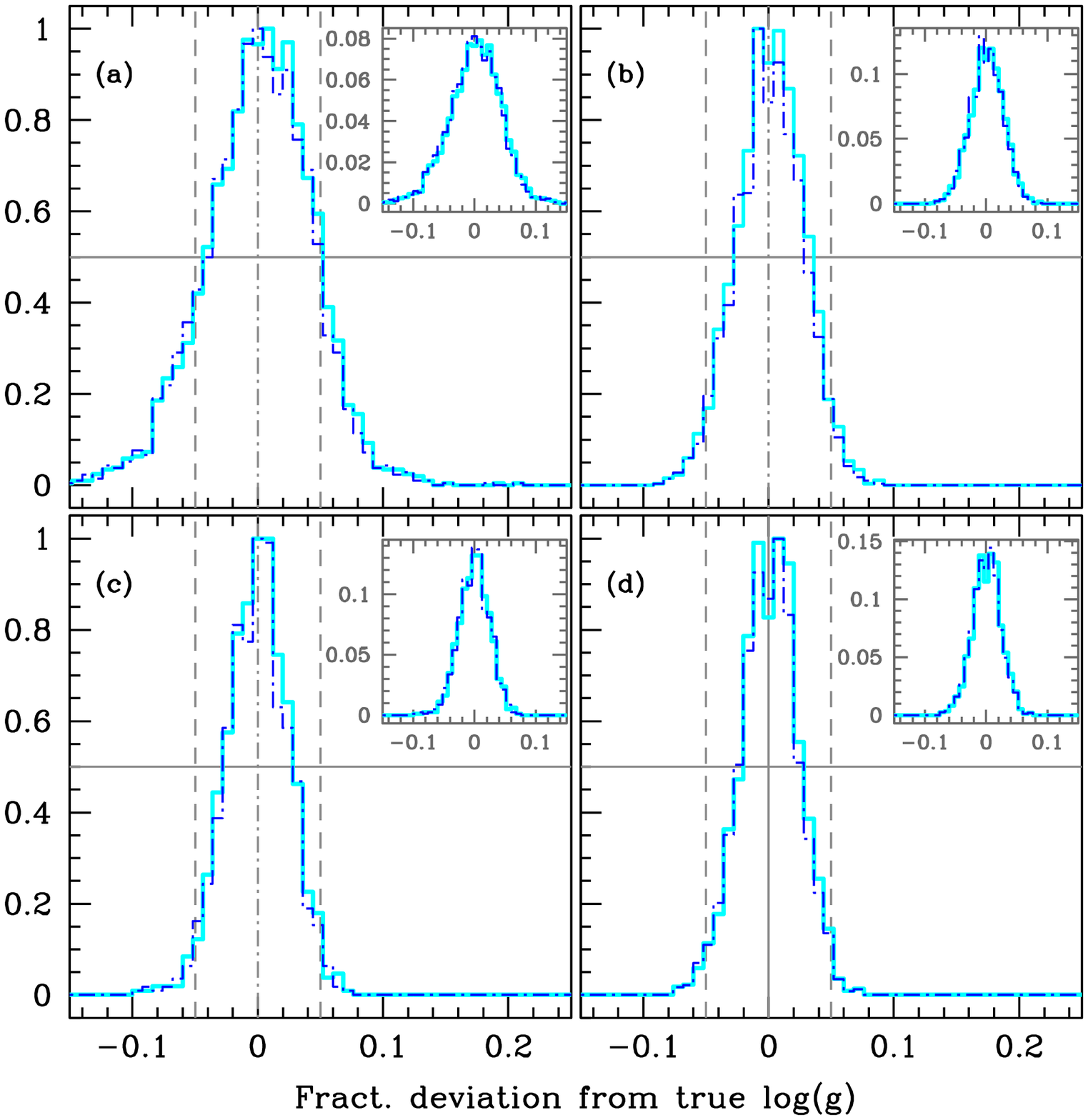} \caption{The
same as Figures~\ref{fig:e_g_yrec_sel_hist}, but
but for stars in selected $\Delta\nu$ ranges. We show results for
stars with $\Delta\nu \le 20\mu$Hz (panel a), $ 20 < \Delta\nu \le
75\mu$Hz (panel b), $75 < \Delta\nu \le 100\mu$Hz (panel c), and $
\Delta\nu >  100\mu$Hz (panel d). We only show the grid results
using the $(\Delta\nu, \nu_{\rm max}, T_{\rm eff})$ combination
(thick cyan line) and the $(\Delta\nu, \nu_{\rm max}, T_{\rm eff},
Z)$ combination (dot-dashed blue line).}
\label{fig:e_g_yrec_sel_hist}
\end{figure}

As with mass and radius, \logg\ is easier to obtain for some stars
than for others. The error histograms for stars with different \dnu\
ranges are shown in Figure~\ref{fig:e_g_yrec_sel_hist}. As can be
seen, errors in \logg\ are smaller for large \dnu\ stars (which are
large \logg\ stars too) than for small \dnu\ stars (which happen to be
small \logg\ stars). However, in all cases the HWHM of the
distributions is less than 5\%.  
 The quantity \logg\ is
difficult to determine precisely using spectroscopy. Errors can be large,
typically 0.1 dex.
For instance for {\it Kepler} star KIC 11026764 different groups, using the
star's spectrum, find \logg\ values ranging from $3.84\pm 0.10$ to $4.19\pm 0.16$
(Metcalfe et al. 2010). The uncertainties in \logg\ from
seismology are thus many  times smaller than those obtained spectroscopically.
Thus it appears that seismology might provide the cleanest way to
determine \logg. 

\subsection{Age}
\label{subsec:age}

Determining stellar ages is crucial for studies of stellar and
galactic evolution.  The way this is normally done is to fit
theoretical evolutionary sequences (e.g. Edvardsson et al.  1992; Ng \& Bertelli
1992; Pont \& Eyers 2004, etc.)  or theoretical isochrones
particularly in the case of star cluster (first attempted by  Demarque \& Larson 1964
and followed by many others).
The approach works
reasonably well for star clusters as long as the color-magnitude
diagram is well defined and stars in all stages of evolution, in
particular, the main-sequence, the main-sequence turnoff and the
red-giant branch, are present. A similar approach is also used for
individual stars, and theoretical tracks are interpolated to the observed
parameters of a given star.  Pont \& Eyer~(2004, 2005) have quantified
some of the biases that hamper determination of stellar
ages. Pont \& Eyer~(2005) claim that the traditional isochrone ages
for field stars are subject to a large systematic bias that they call
`terminal age bias', which tends to pull all ages towards the
end-of-main-sequence lifetime. This is believed to be a result of the
interaction of the observational uncertainties with the strongly
varying speed of evolution of stars in the temperature-luminosity
diagram. They claim that more sophisticated statistical treatments,
particularly Bayesian estimates, are needed to get unbiased
ages. J{\o}rgensen \& Lindegren (2005) and Takeda et al. (2007) also
developed Bayesian-based approaches to determine the ages of
individual field stars. Of these, J{\o}rgensen \& Lindegren's approach
is very similar to our grid approach, especially since the likelihood
function we define can be modified to include any prior.

The two easily observed asteroseismic quantities, \dnu\ and \numax, do
not contain any explicit dependence on age. Age estimates using these
two quantities thus rely on what models predict about how \dnu\ and
\numax\ change with age. And hence, unlike radius and \logg, we know
from the very outset that we cannot get model-independent age
estimates. The question is how large the model dependence is, and
whether we can do any better than non-seismic estimates. Non-seismic
estimates rely on measurements of temperature, metallicity and
luminosity. Most field stars do not have distance measurements and
hence luminosity is difficult to determine. Once we have \dnu\ and
\numax, we do not need to know luminosity since the two seismic
quantities contain mass and radius. Prior knowledge of luminosity
would, of course, make the age estimates more precise.

 It should be noted that model-dependent ages of individual stars can be determined to
extremely high precision once frequencies of a reasonable number of
individual modes are known (see e.g., Metcalfe et al. 2010). We will
not be able to get robust estimates of individual frequencies for all
the {\it Kepler} stars, hence it is important to be able to test
determination of ages using \dnu\ and \numax. It should also be noted
that precise (but again model dependent) ages of main sequence stars
can be determined if the so-called small frequency separation
$\delta\nu$ is known.  This quantity varies with the central hydrogen
abundance, and hence age, of a star. Christensen-Dalsgaard (1988)
suggested that a plot of $\delta\nu$ against \dnu\ can be used to
determine stellar ages. Modifications to this so-called `JCD diagram'
have been suggested by Mazumdar~(2005) and Tang et
al. (2008). However, this method does not work for more evolved stars
(White et al., in preparation). We therefore rely just on \dnu\ and
\numax\ for this analysis.

Since Equations~(\ref{eq:delnu}) and (\ref{eq:numax}) do not contain
time explicitly, there is, as noted above, no direct method to
determine age. Our preliminary investigations with the grid method
have shown that unlike the cases of radius, \logg, and mass, we cannot
get any estimate of age using \dnu\ and \numax\ alone. Hence we try
only two sets of inputs --- (\dnu, \numax, \teff) and (\dnu, \numax,
\teff, [Fe/H]).

\begin{figure}
\epsscale{0.90} \plotone{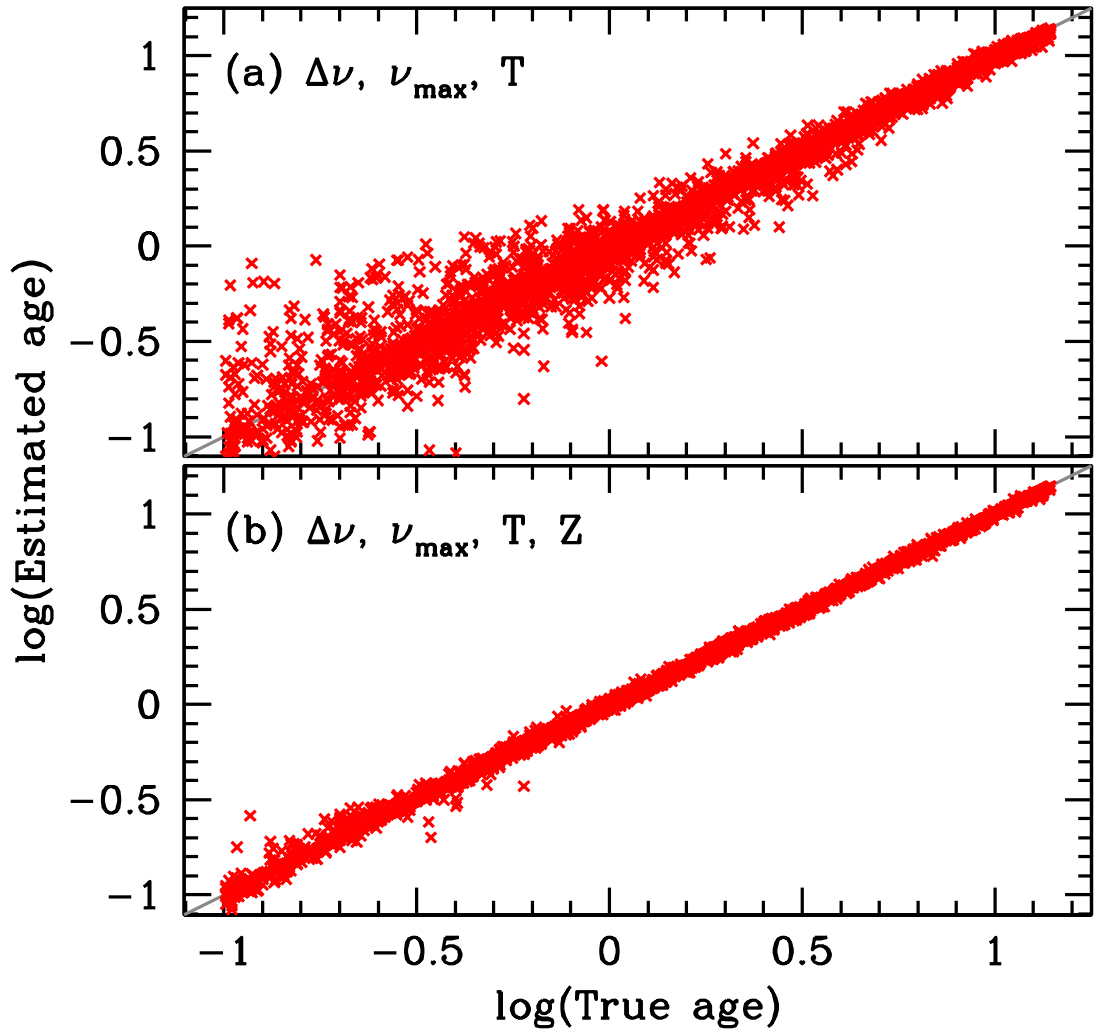} \caption{Age estimates (in Gyr)
obtained by the grid method using two different data combinations.
No errors were added to the data. Both the ``stars'' and the grid
were based on YREC models. } \label{fig:ne_age_y_grid}
\end{figure}

\begin{figure}
\epsscale{0.6} \plotone{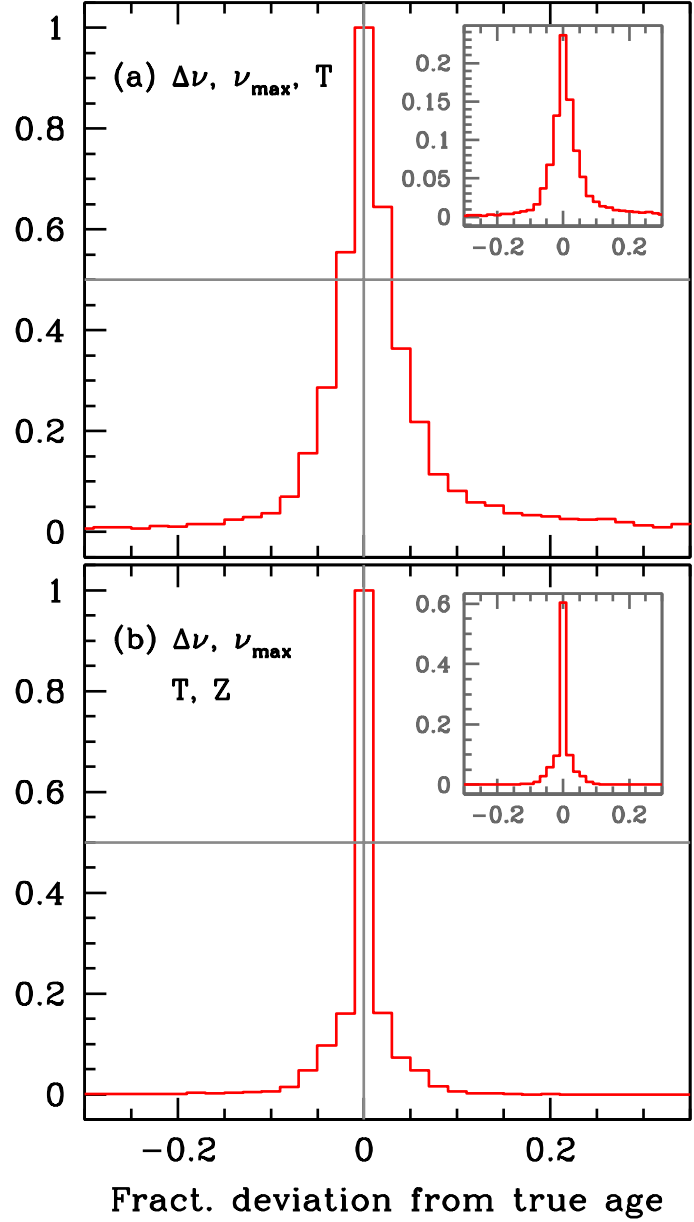} \caption{Histograms
showing the fractional deviation between the true ages and the ages
estimated by the grid method using YREC models
to illustrate the accuracy of the determinations for error-free
data. The horizontal gray line is the half-maximum, and the vertical
gray line indicates accurate results.
 As with other figures, the inset in each panel
shows the distributions normalized to unit area.}
 \label{fig:ne_age_yrec_hist}
\end{figure}

\begin{figure}
\epsscale{0.6} \plotone{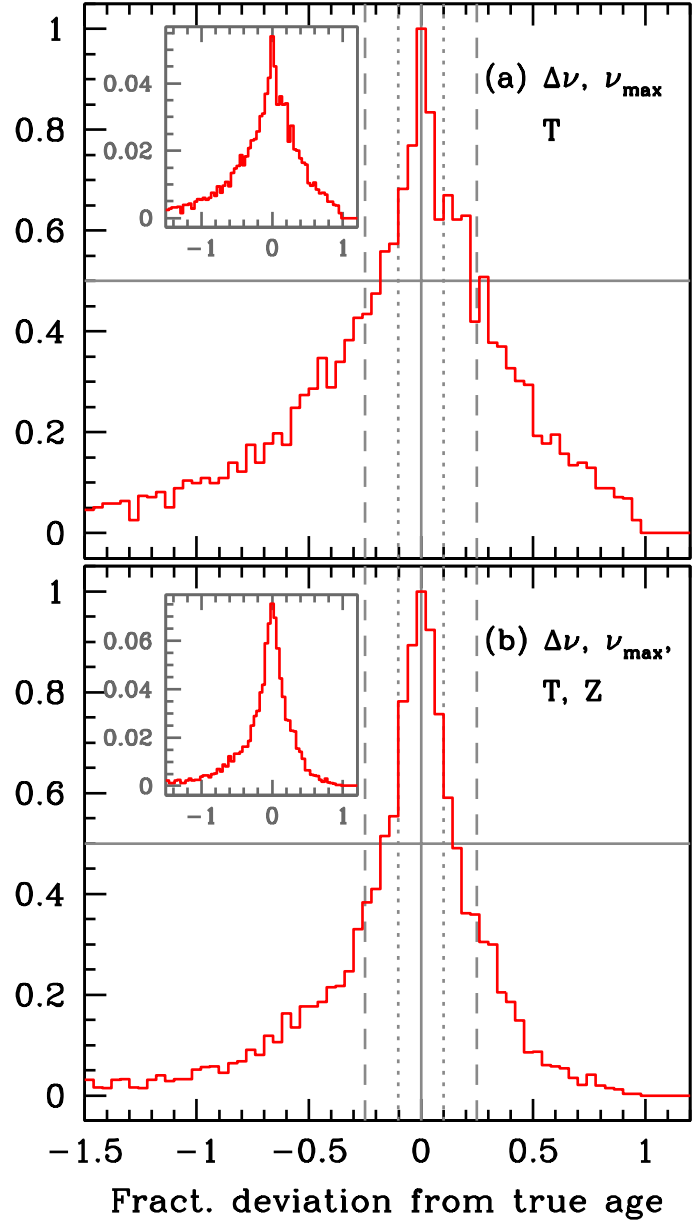} \caption{The
same as Figure~\ref{fig:ne_age_yrec_hist} but for the case when
errors were added to data. The dashed gray line shows $\pm 25$\%
error in results, the dotted gray lines mark $\pm10$\% error. }
\label{fig:e_age_yrec_s_hist}
\end{figure}

Figure~\ref{fig:ne_age_y_grid} shows the result of estimating the ages
of about 7300 stars with error-free data. The stars are based on YREC
models, and the results were obtained using the YREC grid. As can be seen, even
with a similar grid and error-free data, knowledge of $Z$ is a
must. This is not surprising, since stars of the same mass evolve at
different rates depending on their metallicity. The error histograms
for these cases are shown in Figure~\ref{fig:ne_age_yrec_hist}. The
figure shows that using metallicity reduces the HWHM of the
distribution from about 5\% to less than 2.5\% and the maximum error
is reduced to 10\%. Adding errors to the inputs of course makes the
situation worse. The error histograms for this case are shown in
Figure~\ref{fig:e_age_yrec_s_hist}. Without metallicity, the HWHM is
about 20\% and the distribution has a large tail. Once metallicity is
used, the HWHM reduces to around 15\%, and while the distribution
still has a long tail, it is not as wide as it was earlier.

\begin{figure}
\epsscale{0.8} \plotone{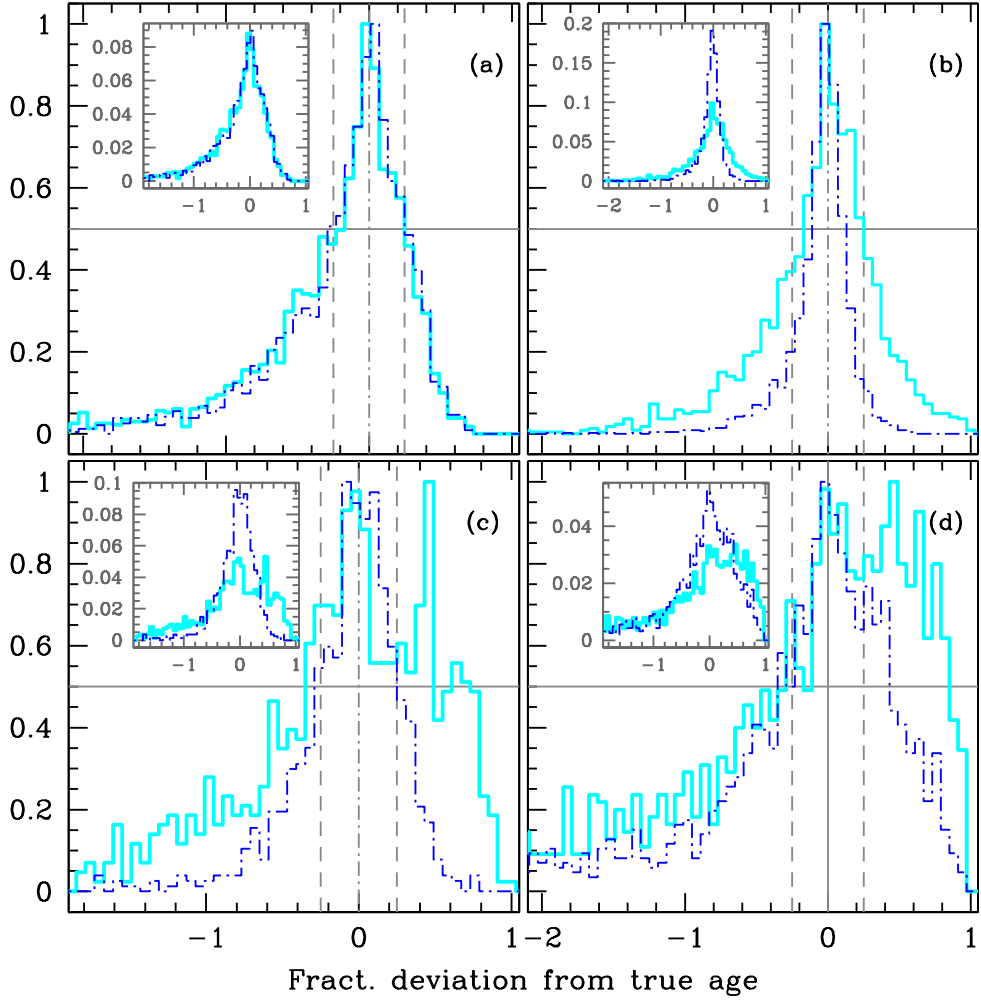} \caption{The
same as Figure~\ref{fig:e_age_yrec_s_hist}, but for stars grouped
into different \dnu\ ranges. The thick cyan line shows the result
of using (\dnu, \numax, \teff) as inputs, the blue dot-dashed lines
are the results of using  (\dnu, \numax, \teff, [Fe/H]) as
inputs. The solid gray vertical line is for zero deviation, while 
the dashed lines show $\pm 25$\% error. We show results for
stars with $\Delta\nu \le 20\mu$Hz (panel a), $ 20 < \Delta\nu \le
75\mu$Hz (panel b), $75 < \Delta\nu \le 100\mu$Hz (panel c), and $
\Delta\nu >  100\mu$Hz (panel d).
 The inset in each panel shows the distributions normalized
to unit area.
}
\label{fig:e_age_yrec_sel_hist}
\end{figure}

As in the cases of radius, mass and \logg, we repeated the analysis
after grouping stars in \dnu\ and the results are shown in
Figure~\ref{fig:e_age_yrec_sel_hist}. It appears that using this
method, we can find ages of sub-giants and red-giants more precisely
than ages of main sequence and turnoff stars. In retrospect, this is
not surprising -- the rapid variation of stellar radius with age for
evolved stars implies a large change in both \dnu\ and \numax\, making
this method more sensitive. Thus for main-sequence and turn-off stars,
we really need the small-separations to get relatively precise
estimates of age. 
The case of core helium burning red-clump stars is 
interesting. Errors in temperature and metallicity usually means that the
grid method include stars from the ascending part of the red-giant
branch in the likelihood function calculations. The converse is also 
true, the likelihood function of stars on the ascending part of the 
red-giant branch can include red-clump stars. This results in
age-errors that are larger than those for subgiant stars.

\begin{figure}
\epsscale{0.9} \plotone{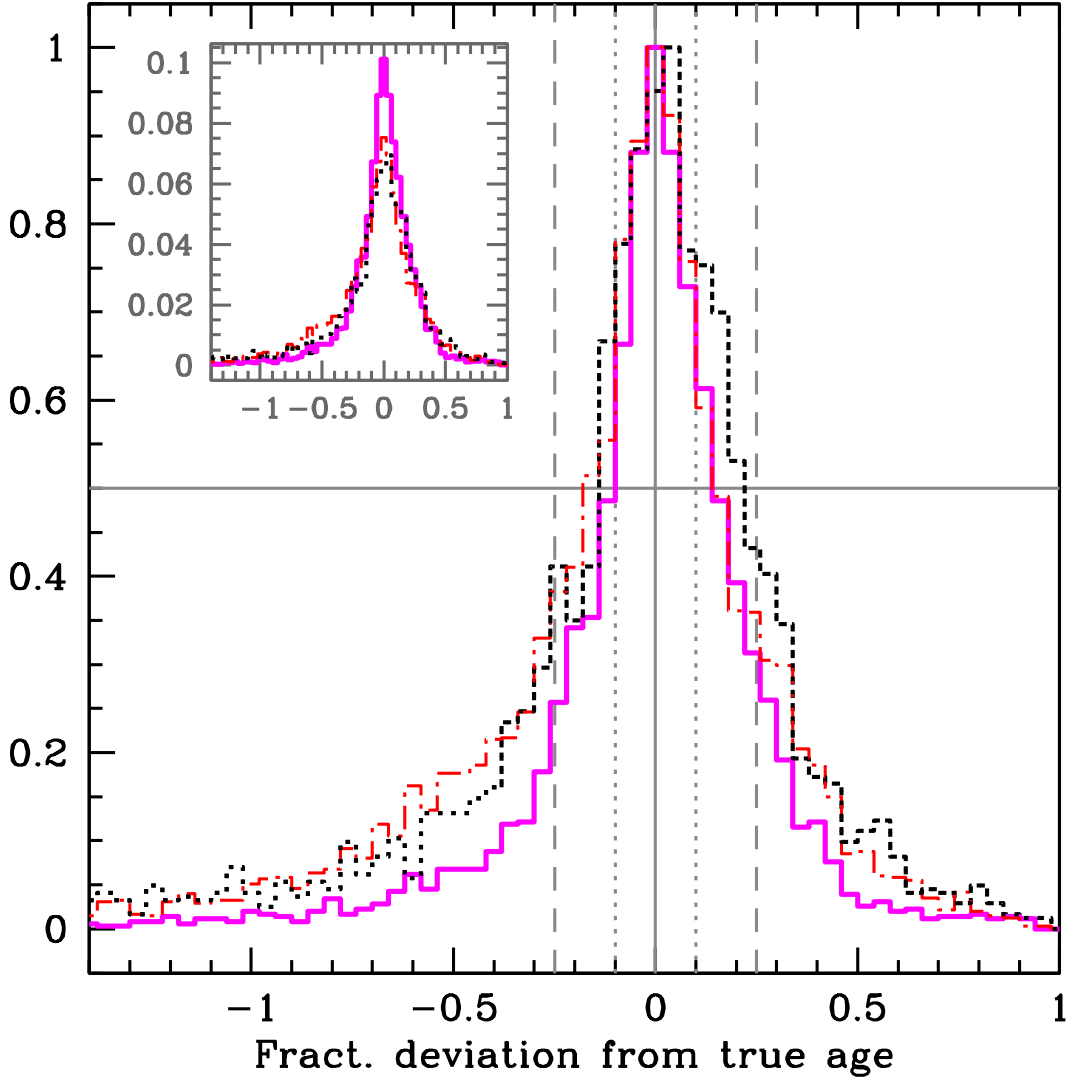}
\caption{A comparison of histograms showing the fractional deviation between
the true ages and the ages obtained using combinations of  seismic
and non-seismic data with that for non-seismic data only. 
The dotted black line is for results obtained
using only the non-seismic parameters $(T_{\rm eff}, Z, M_v)$, red
dot-dashed for $(\Delta\nu, \nu_{\rm max}, T_{\rm eff}, Z)$ and the
magenta solid line for $(\Delta\nu, \nu_{\rm max}, T_{\rm eff}, Z,
M_v)$. The dashed gray line shows $\pm 25$\% error in results, the
dotted gray lines mark $\pm10$\% error. 
 The inset shows the distributions normalized to unit
area.
}
\label{fig:e_age_yrec_all_hist}
\end{figure}

An important question is whether using seismic data actually helps us
at all. In Figure~\ref{fig:e_age_yrec_all_hist} we show the error
histogram for the seismic case with inputs (\dnu, \numax, \teff,
[Fe/H]) and compare it with that of a non-seismic case with inputs
(\teff, [Fe/H], and $M_V$), where we have assumed errors of
0.1\,mag in $M_V$. The figure confirms our earlier assertion
that the information contained in \dnu\ and \numax\ allows us to get
comparable error estimates without knowing the luminosity of the
stars. If we knew luminosity as well as \dnu\ and \numax, we would do
much better, as we also show in Figure~\ref{fig:e_age_yrec_all_hist}.

\begin{figure}
\epsscale{0.6} \plotone{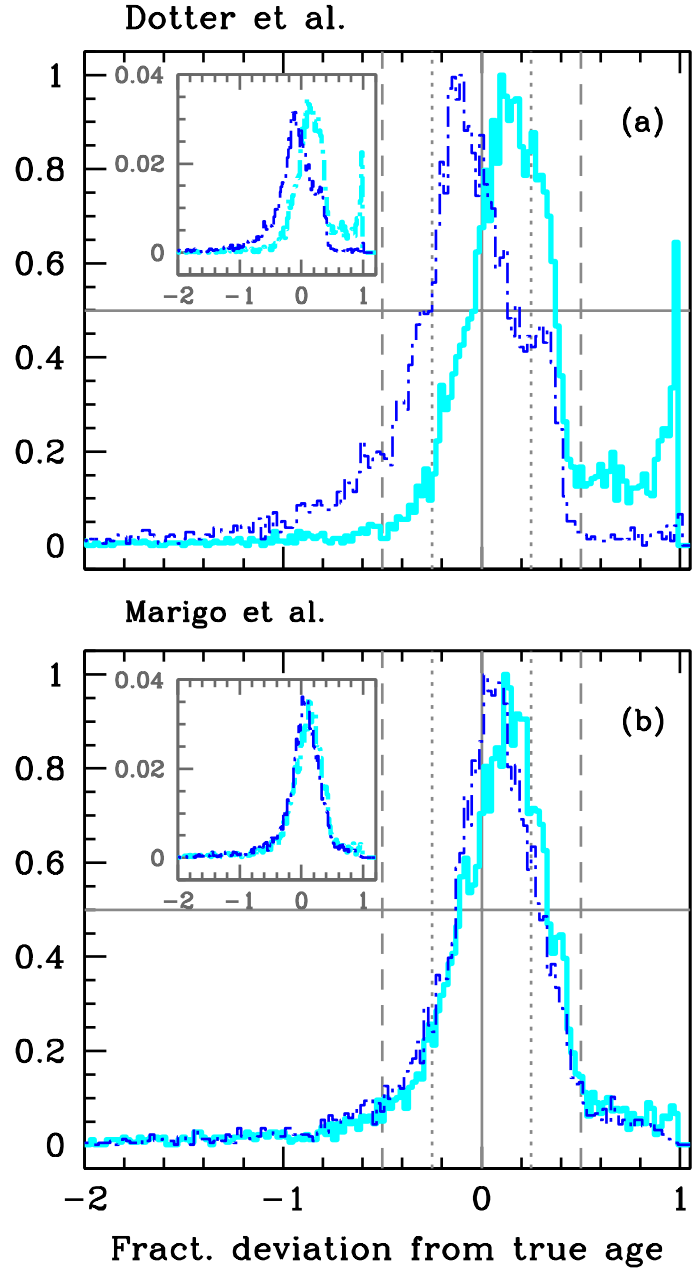} \caption{ The same
as Figure~\ref{fig:ne_age_yrec_hist} but when a different grid of
models was used. Panel (a) shows the results of using the Dotter et
al. grid, and panel (b) shows results for Marigo et al. grid. The
thick cyan lines are results of using the $(\Delta\nu, \nu_{\rm
max}, T_{\rm eff})$ combination, and the dot-dashed blue line is for
the $(\Delta\nu, \nu_{\rm max}, T_{\rm eff}, Z)$ combination.}
\label{fig:ne_age_other_hist}
\end{figure}

\begin{figure}
\epsscale{0.6} \plotone{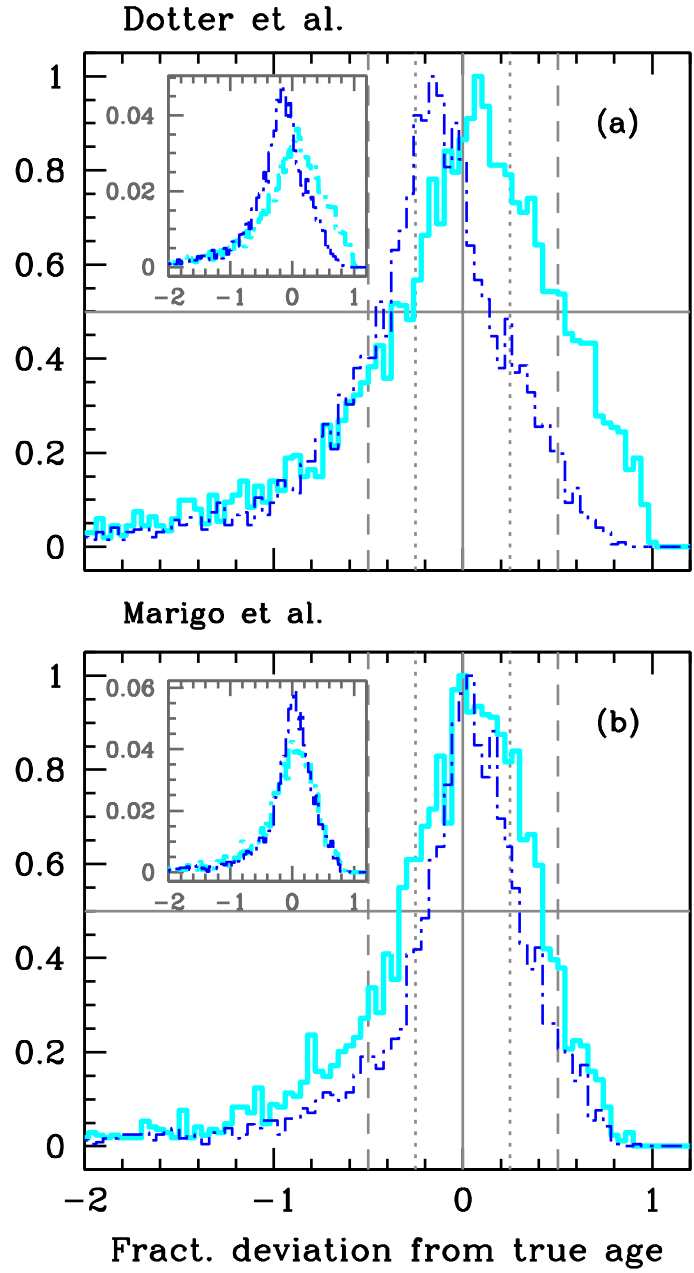} \caption{ The same as
Figure~\ref{fig:ne_age_other_hist} but for the case with errors
added to the data. } \label{fig:e_age_other_hist}
\end{figure}

All the age results are, of course, model dependent.  This is shown in
\ref{fig:ne_age_other_hist} for the error-free case where we plot the
error histograms for ages obtained with the Marigo et al. and Dotter
et al. grid. As expected, the error is larger, and the case for
needing metallicity is stronger, with the HWHM being about 20\%. In
Figure~\ref{fig:e_age_other_hist} we show what happens when errors are
added. As with the other global parameters, it appears that input
errors can dominate over the systematic errors. And the figures drive
home the need for good metallicity measurements to derive ages
properly.  The model dependence and uncertainties of the method
are demonstrated using solar data. Estimates of the solar age are
listed in Table~\ref{tab:age}. We can see that unless we use metallicity, the
error-bars are large enough to make the results essentially useless. 
Even when we use metallicity, the errors are large, though the central value is more acceptable.

\begin{table}
\caption{The solar age obtained with BiSON data using different grids and
input combinations.}
\begin{center}
\begin{tabular}{lcc}
\hline
\noalign{\smallskip}
Grid  & \multispan2{\hfill Input combinations\hfill}\\
 & (\dnu, \numax, \teff) & (\dnu, \numax, \teff, $Z$) \\
\noalign{\smallskip}
\hline
\noalign{\smallskip}
Dotter et al. & $2.96^{+5.57}_{-1.15}$ & $4.62^{+2.13}_{-1.73}$ \\
Marigo et al. & $3.98^{+6.02}_{-3.63}$ & $3.55^{+2.80}_{-2.75}$ \\
YREC &          $5.57^{+6.04}_{-5.05}$ & $3.97^{+2.07}_{-1.84}$ \\
\noalign{\smallskip}
\hline
\end{tabular}
\end{center}
\label{tab:age}
\end{table}

\begin{figure*}
\epsscale{0.5}
\centerline{\plotone{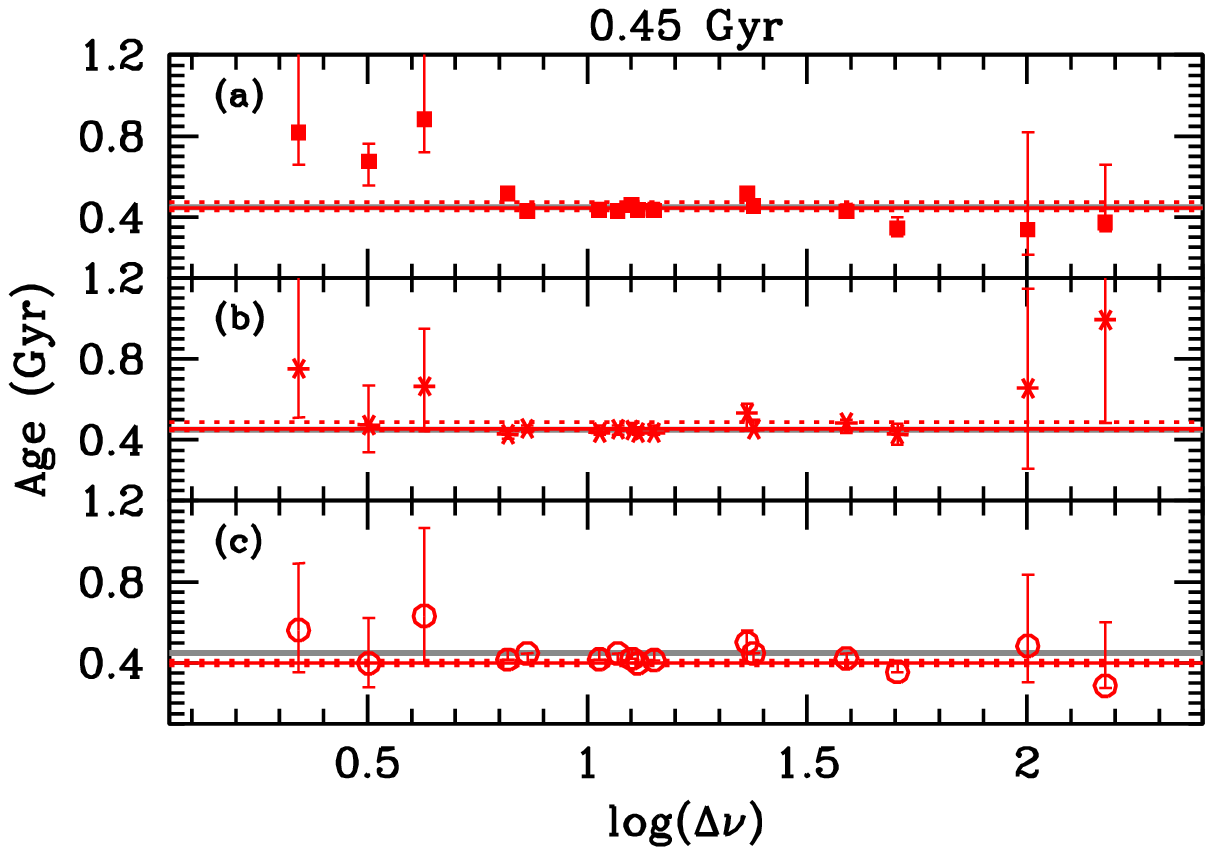}\plotone{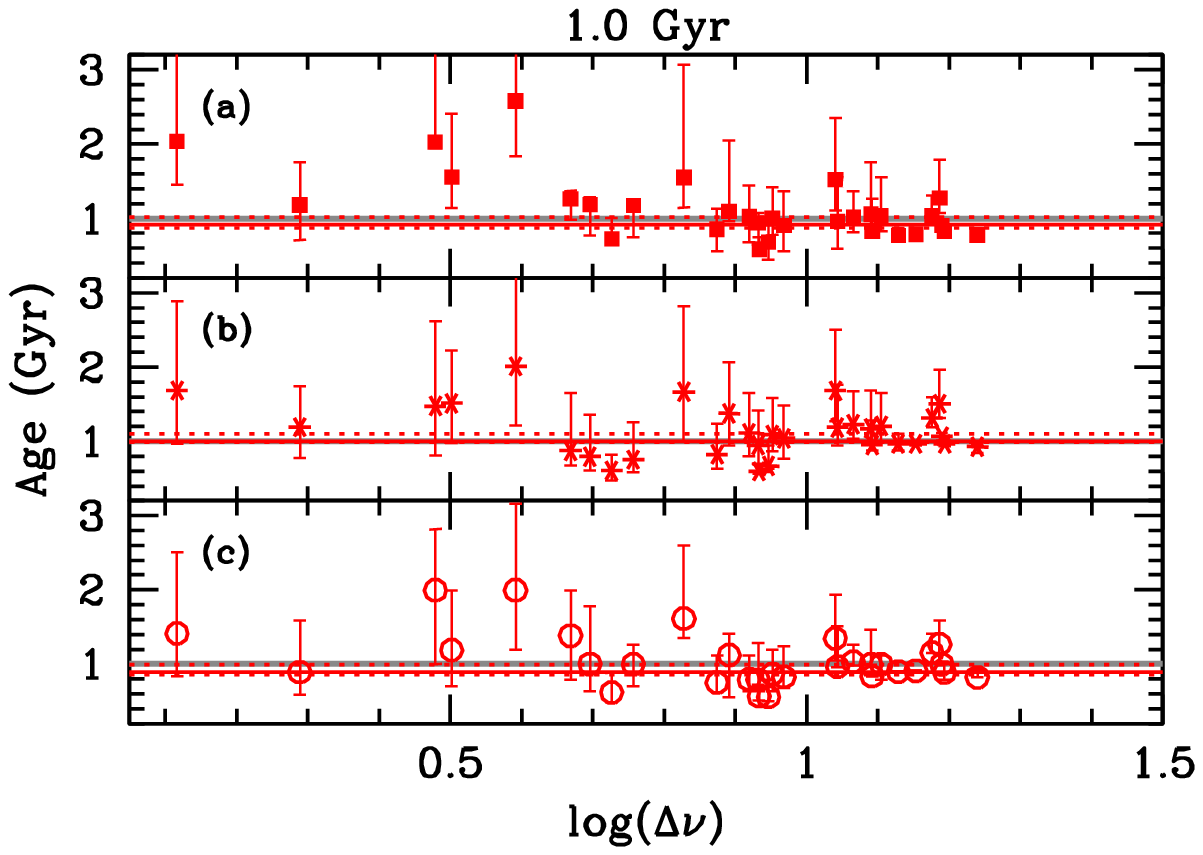}}
\centerline{\plotone{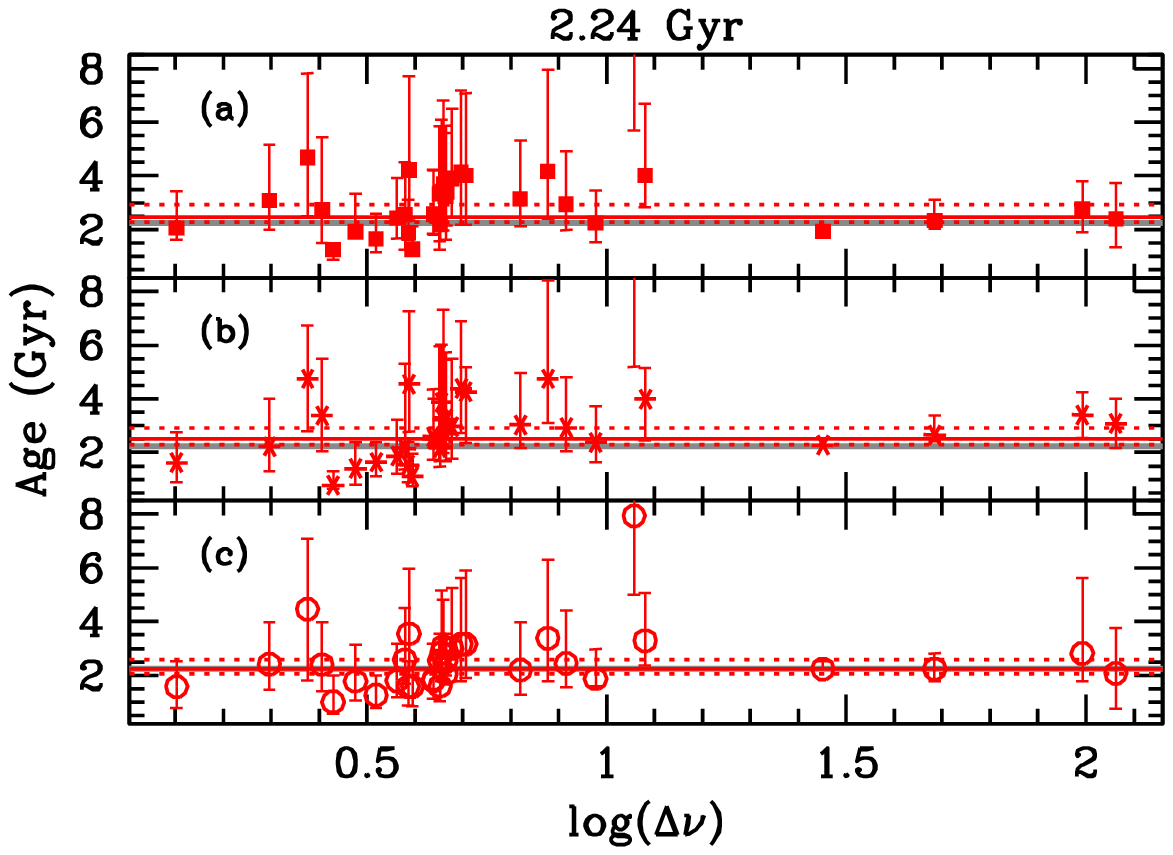}\plotone{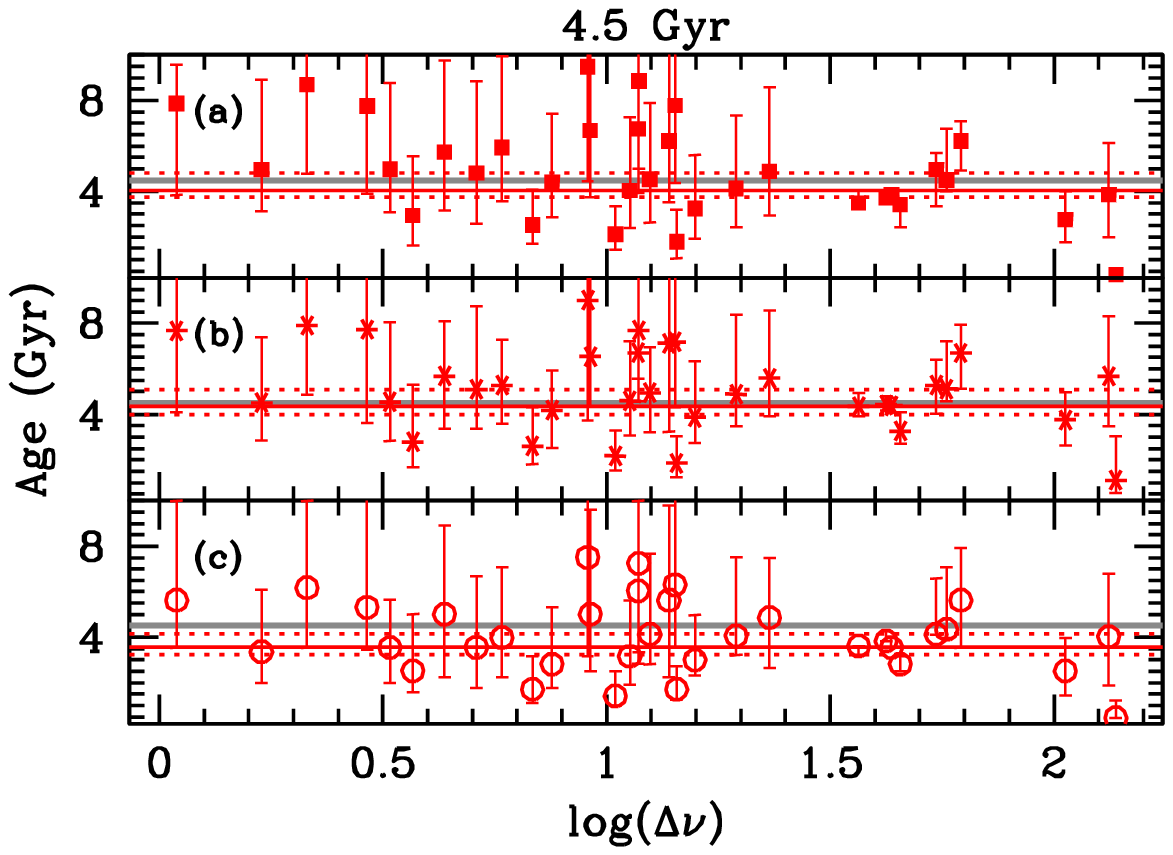}}
\caption{Age results obtained for star clusters. Each main panel
shows the result from a simulated cluster of a given age. Sub-panel (a) shows
results obtained with the YREC grid, (b) with the Dotter et al. grid
and (c) with the Marigo et al. grid. In each sub-panel, the
horizontal line marks the true age of the cluster, the points with the
error bars are the results for each individual star. The red solid
line is the result obtained assuming that all stars in the cluster
have the same age, while the dotted lines show the $1\sigma$ error
bars on that result.} \label{fig:cluster}
\end{figure*}

The {\it Kepler} field-of-view contains four open clusters, including
NGC~6791 and NGC~6819. Thus the question arises whether we can apply
the grid method to determine ages of clusters with the handful of
stars that are expected to show detections of solar-like
oscillations. Clusters provide two advantages: we can apply the prior
that all stars have the same age; and, given the intrinsic interest in
clusters, they usually have good metallicity estimates.
To determine cluster ages, we first determine the ages of the
individual stars, which allows us to remove outliers, and then we
re-derive the age of the cluster as a whole after applying the prior
that the stars have the same age. The results for four simulated
clusters are shown in Figure~\ref{fig:cluster}. Two of the clusters
were derived from the Dotter et al. grid and two from the Marigo et
al.  grid, and we derive the ages using all three grids. For each we
have between 20 and 30 stars, which is what we expect for initial {\it
Kepler} data on clusters. We used stars that populate all parts of the
color-magnitude diagram, from the main sequence to the red-clump. As
can be seen from the figure, it is possible to get precise results for
clusters, even with a handful of stars.  There are systematic errors
though, that result from the differences in the physics and more importantly
composition differences between the proxy stars and the grid used. Thus it 
is extremely important to that before determining ages we first construct
a grid of models that correspond to the composition of the cluster.

\begin{figure*}
\epsscale{0.5} \centerline{\plotone{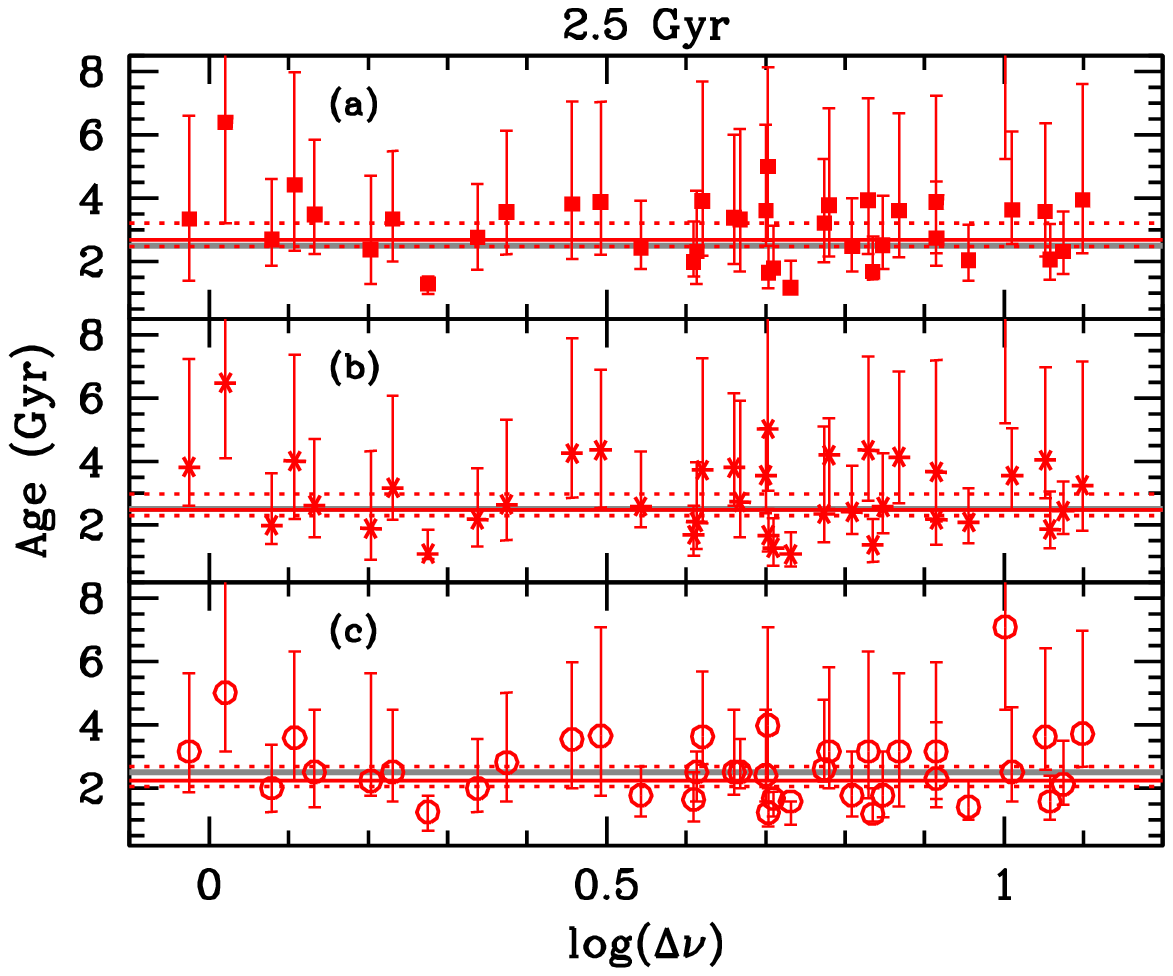}}
\centerline{\plotone{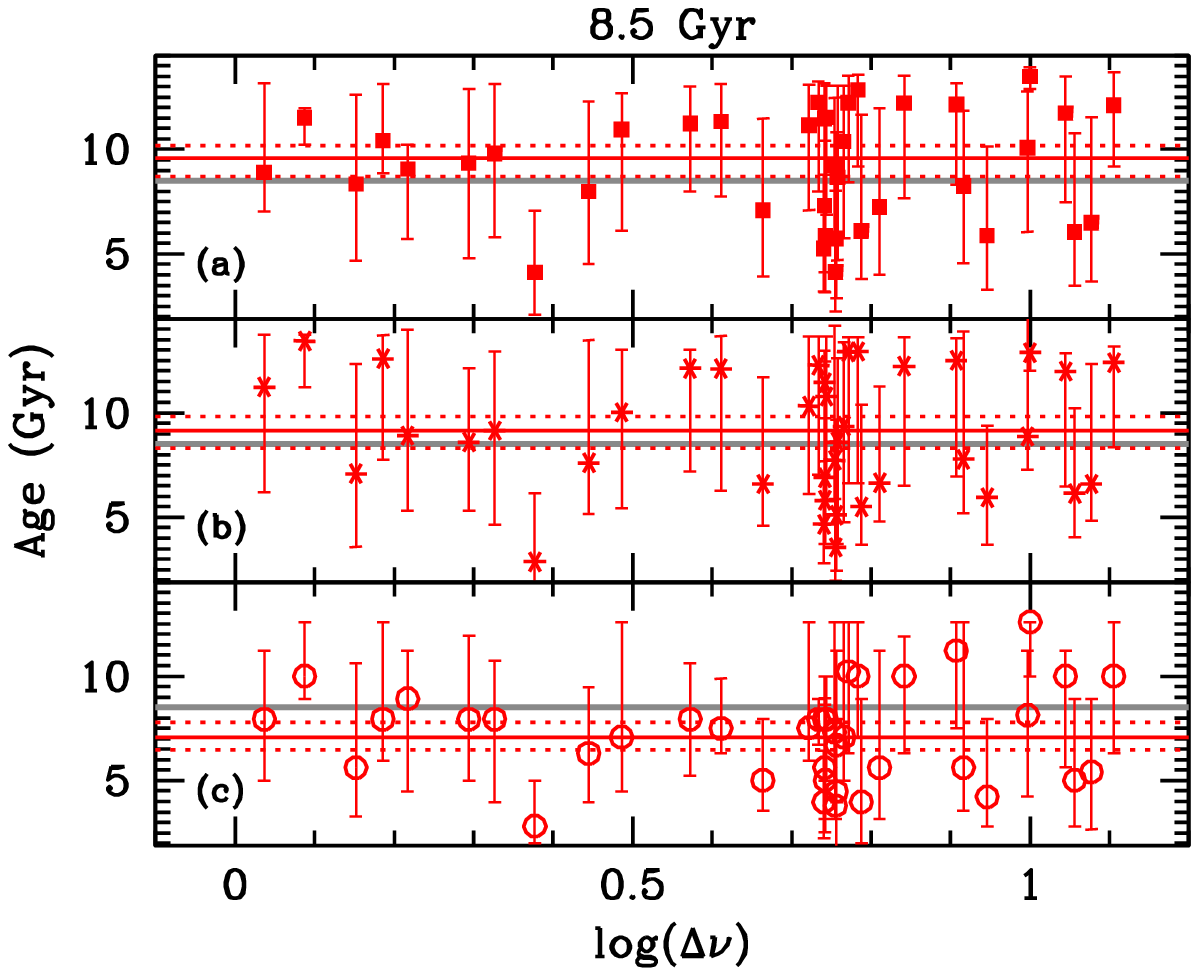}} \caption{ Same as
Figure~\ref{fig:cluster}, but for the case when only data on red
giants and red-clump stars are available. The upper panel has stars
of an age similar to that of NGC 6819, the lower panel has stars
with age similar to NGC 6791.} \label{fig:ngc}
\end{figure*}

Initial data for NGC 6791 and NGC 6819 will only be for red-giants and
core helium burning red-clump stars. We know from Figure~\ref{fig:e_age_yrec_sel_hist}
that having only giants will not be a handicap. However, we test this
by simulating a cluster of age 2.5 Gyr (roughly the age of NGC~6891)
and one of 8.5 Gyr (around the age of NGC~6791) and use only red-giant
and clump stars. The results are shown in Figure~\ref{fig:ngc}.  As
can be seen, we expect precise results for both clusters (with the
older cluster having a larger systematic error). It should be noted
that these results are independent of the distance-modulus of the
clusters.  They will depend on the extinction, through temperatures.
In addition to ages, we can give model-independent estimates of the
distances to the clusters by determining the radii and masses from the
seismic data.

\section{Conclusions}
\label{sec:conc}

We made an in-depth study of grid-based asteroseismic analysis to
determine possible systematic and other errors in radius and mass
estimates derived using seismic data.

We find that when errors in the seismic parameters and \teff\ are included,
 error estimates in radius and mass are higher when
determined directly from the equations defining the seismic quantities
\dnu\ and \numax\ than when determined using the grid method.  This is
not surprising. Equations~(\ref{eq:delnu}) and (\ref{eq:numax}) assume
that all values of \teff\ are possible for a star of a given mass and
radius. However, the equations of stellar structure and evolution tell
us otherwise --- we know that for a given mass and radius, only a
narrow range of temperatures are allowed. The grid method takes this
into account implicitly since the grid is constructed by solving the
equations of stellar structure and evolution.

However, grid based asteroseismology can lead to some model
dependence. While there is almost no model dependence in derived
values of radius, there can be considerable model dependence in mass,
unless we have reasonable measures of metallicity. However, given the
expected errors in the seismic and non-seismic inputs, the systematic
errors are much smaller than the error in the results caused by errors
in the observations.

Errors in mass and radius estimates are positively correlated and the
correlation is high. The correlation is such that it leads to very
small errors in $M/R^2$ and hence \logg. We find that we can determine
\logg\ precisely and accurately with both direct and grid-based methods
and that there is no systematic error in the estimates. Given that
\logg\ estimates from spectroscopy can be extremely imprecise,
seismology gives us an alternative to spectroscopy when it comes to
determining \logg.

Given the near model-independence of the radius and \logg\ estimates,
it may be tempting to assume that we could get near-model independent
mass estimates from estimated values of radius and \logg. This cannot
be done to high precision because although the errors in radius and
\logg\ are small, they get magnified when the error on mass is
calculated. Since $M=gR^2$, the relative error in mass is, at the very
least, the relative error in $g$ and twice the relative error in $R$,
added in quadrature. There is a third term that arises from the
correlation between $g$ and $R$, which adds to the error in mass. This
is not a surprising result since whether we estimate mass on its own
or from estimates of radius and \logg, the information available is
exactly the same.

As far as ages are concerned, seismic data do not give us any direct
information, however, since the radius and mass of a star are a
function of age, seismic data merely give us two extra pieces of
information. Since neither \dnu\ nor \numax\ are explicit functions of
age, we can only get model-dependent age estimates. The advantage that
seismic data give us is that we do not need to know the distance and
absolute luminosity of the star to determine the age through
models. The model dependence in the results can be minimized if the
abundances of the stars are known. The errors can be rather large,
easily as large as 25\%. We find however, that we can determine the
ages of star clusters to a much higher precision when we apply the
prior that all cluster stars have the same age. And unlike isochrone
fitting techniques, asteroseismic cluster ages are independent of the
distance modulus and can be determined with only a handful of
stars. Cluster ages can be determined even if we use only red-giant
stars, which is encouraging since initial seismic data from clusters
in the {\it Kepler} field of view will only be data on the cluster
red-giants.

\acknowledgments

N.G. acknowledges the China State Scholarship Fund that allowed her to
spend a year at Yale. She also acknowledges grant 2007CB815406 of the
Ministry of Science and Technology of the Peoples Republic of China
and grants 10773003 and 10933002 from the National Natural Science
Science Foundation of China. W.J.C. and Y.E. acknowledge the
financial support of the UK Science and Technology Facilities Council
(STFC), and the International Space Science Institute (ISSI).

\newpage
\clearpage
\newpage

\newpage
\clearpage

\newpage
\clearpage

\newpage
\clearpage

\newpage
\clearpage

\newpage
\clearpage

\newpage
\clearpage

\newpage
\clearpage

\newpage
\clearpage

\newpage
\clearpage

\newpage
\clearpage

\newpage
\clearpage

\newpage
\clearpage

\newpage
\clearpage

\newpage
\clearpage

\newpage
\clearpage

\newpage
\clearpage

\newpage
\clearpage

\newpage
\clearpage

\newpage
\clearpage

\newpage
\clearpage

\newpage
\clearpage

\newpage
\clearpage

\newpage
\clearpage

\newpage
\clearpage

\newpage
\clearpage

\newpage
\clearpage

\newpage
\clearpage
\end{document}